\documentclass[twocolumn]{aastex631}

\usepackage{booktabs}
\usepackage{makecell}
\usepackage{gensymb}
\usepackage{threeparttable}

\usepackage{ulem}

\usepackage{soul} 

\newcommand{\s}{\,{\rm s}}      
\newcommand{\ps}{\,{\rm s}^{-1}}

\newcommand{\km}{\,{\rm km}}
\newcommand{\kms}{$\km\ps$}

\newcommand{\kpc}{\,{\rm kpc}}

\newcommand{\K}{\,{\rm K}}

\newcommand{\gray}{{\rm $\gamma$-ray}}
\newcommand{\grays}{{\rm $\gamma$-rays}}

\newcommand{\snr}{G35.6$-$0.4}
\newcommand{\seast}{G35.6$-$0.5}
\newcommand{\hess}{{HESS~J1858$+$020}}
\newcommand{\lhaaso}{{1LHAASO~J1857$+$0203u}}

\newcommand{\twCO}{$^{12}$CO}   
\newcommand{\thCO}{$^{13}$CO}

\newcommand{\Jotz}{$J$=1--0}

\newcommand{\Ep}{E_{\rm p}}
\newcommand{\Ee}{E_{\rm e}}
\newcommand{\Eb}{E_{\rm b}}

\begin{document}

\title{An Enigmatic PeVatron in an Area around HII Region \seast}

\author{Zhen Cao}
\affiliation{Key Laboratory of Particle Astrophysics \& Experimental Physics Division \& Computing Center, Institute of High Energy Physics, Chinese Academy of Sciences, 100049 Beijing, China}
\affiliation{University of Chinese Academy of Sciences, 100049 Beijing, China}
\affiliation{TIANFU Cosmic Ray Research Center, Chengdu, Sichuan,  China}
 
\author{F. Aharonian}
\affiliation{Dublin Institute for Advanced Studies, 31 Fitzwilliam Place, 2 Dublin, Ireland }
\affiliation{Max-Planck-Institut for Nuclear Physics, P.O. Box 103980, 69029  Heidelberg, Germany}
 
\author{Axikegu}
\affiliation{School of Physical Science and Technology \&  School of Information Science and Technology, Southwest Jiaotong University, 610031 Chengdu, Sichuan, China}
 
\author{Y.X. Bai}
\affiliation{Key Laboratory of Particle Astrophysics \& Experimental Physics Division \& Computing Center, Institute of High Energy Physics, Chinese Academy of Sciences, 100049 Beijing, China}
\affiliation{TIANFU Cosmic Ray Research Center, Chengdu, Sichuan,  China}
 
\author{Y.W. Bao}
\affiliation{School of Astronomy and Space Science, Nanjing University, 210023 Nanjing, Jiangsu, China}
 
\author{D. Bastieri}
\affiliation{Center for Astrophysics, Guangzhou University, 510006 Guangzhou, Guangdong, China}
 
\author{X.J. Bi}
\affiliation{Key Laboratory of Particle Astrophysics \& Experimental Physics Division \& Computing Center, Institute of High Energy Physics, Chinese Academy of Sciences, 100049 Beijing, China}
\affiliation{University of Chinese Academy of Sciences, 100049 Beijing, China}
\affiliation{TIANFU Cosmic Ray Research Center, Chengdu, Sichuan,  China}
 
\author{Y.J. Bi}
\affiliation{Key Laboratory of Particle Astrophysics \& Experimental Physics Division \& Computing Center, Institute of High Energy Physics, Chinese Academy of Sciences, 100049 Beijing, China}
\affiliation{TIANFU Cosmic Ray Research Center, Chengdu, Sichuan,  China}
 
\author{W. Bian}
\affiliation{Tsung-Dao Lee Institute \& School of Physics and Astronomy, Shanghai Jiao Tong University, 200240 Shanghai, China}
 
\author{A.V. Bukevich}
\affiliation{Institute for Nuclear Research of Russian Academy of Sciences, 117312 Moscow, Russia}
 
% ...
%\email{}




\begin{abstract}

Identifying Galactic PeVatrons (PeV particle accelerators) from the ultra-high-energy (UHE, $>$100\,TeV) \gray\ sources plays a crucial role in revealing the origin of Galactic cosmic rays.
The UHE source \lhaaso\ is suggested to be associated with \hess, which may be attributed to the possible PeVatron candidate supernova remnant (SNR) \snr\ or HII region \seast.
We perform detailed analysis on the very-high-energy and UHE \gray\ emissions towards this region with data from the Large High Altitude Air Shower Observatory (LHAASO).
1LHAASO J1857$+$0203u is detected with a significance of $11.6\sigma$ above 100\,TeV, indicating the presence of a PeVatron.
It has an extension of $\sim 0.18\degree$ with a power-law (PL) spectral index of $\sim$2.5 in 1--25\,TeV and a point-like emission with a PL spectral index of $\sim$3.2 above 25\,TeV.
Using the archival CO and HI data, we identify some molecular and atomic clouds that may be associated with the TeV \gray\ emissions.
%We use the X-ray and GeV \gray\ results from previous studies to model the broadband spectral energy distribution.
Our modelling indicates that the TeV \gray\ emissions are unlikely to arise from the clouds illuminated by the protons that escaped from SNR \snr.
In the scenario that HII region \seast\ could accelerate particles to the UHE band, the observed GeV-TeV \gray\ emission could be well explained by a hadronic model with a PL spectral index of $\sim$2.0 and cutoff energy of $\sim$450\,TeV.
However, an evolved pulsar wind nebula origin cannot be ruled out.

\end{abstract}


\section{Introduction}\label{sect:introduction}
The origin of cosmic rays (CRs) has remained elusive for more than one hundred years since their discovery.
According to \gray\ observations of the Galactic disk and the Magellanic clouds, it is suggested that the CRs below the so-called CR `knee' ($\sim3\times10^{15}$ eV) are produced in our Galaxy \citep[e.g.][]{baade1934,ginzburg1964}.
This indicates that particles can be accelerated up to peta-electron volt energies by some Galactic sources,
called PeV particle accelerators or `PeVatrons'.
The first strong evidence came from observations on the Galactic Centre by H.E.S.S.\citep{hess_GC_2016}, MAGIC\citep{magic_GC_2020}, and VERITAS \citep{veritas_GC_2021}, followed by more observations from HAWC on a few objects in the Galactic plane \citep{hawc2020ApJ,hawc2020prl,hawc2021.cygnus}.
The discovery of dozens of sources by the Large High Altitude Air Shower Observatory (LHAASO) at ultra-high energy (UHE) above 100\,TeV further confirms the existence of Galactic PeVatrons \citep{cao21nat,lh_cat23}.
Among the prominent PeVatron candidates, such as supernova remnants (SNRs), pulsars (PSRs) and their wind nebulae (PWNe), young massive star clusters or HII regions (HIIRs), etc \citep{chen22cpc}, the dominance of PWNe has been established and explicitly reported by both LHAASO \citep[e.g.][]{cao21nat} and HAWC \citep[e.g.][]{hawc_jpcs_2023}.
However, it is unknown which types of objects are hadronic PeVatrons.
Studying the \grays\ produced by the CRs interacting with dense matter surrounding the acceleration site is an important and effective way to unveil the origin of Galactic CRs.

Radio complex \snr\ is likely associated with the unidentified {extended} TeV source \hess\ \citep{hess08,hess18} and was suggested to be a {hadronic} PeVatron candidate \citep{zhang22}.
With the Giant Metrewave Radio Telescope (GMRT) observations at 610\,MHz \citep{paredes14}, the radio complex was resolved into two nearly circularly shaped extended sources, SNR \snr\ and HIIR \seast.
No diffuse X-ray emission but seven X-ray point sources (labelled as X1--7) were detected with a 29.8\,ks \textit{Chandra} observation in the region of \hess\ \citep{paredes14}.
Among these sources,
X1--4 might be embedded protostars projectively distributed on the shell of the HIIR, while X5 might be coincident with a star formation region and close to the southern molecular clump (see \citealt{paredes14}, Fig. 2).
At GeV energies, 
the \textit{Fermi} Large Area Telescope (LAT) Fourth Source Catalog Data Release 4 \citep[4FGL-DR4,][]{fermi_dr3,fermi_dr4} reported two point sources, 4FGL
J1857.6$+$0212 and 4FGL J1858.3$+$0209, towards the G35.6$-$0.4 region.
Of these two GeV sources, 4FGL J1858.3$+$0209 was re-localised as SrcX2 by \citet{cui21} or SrcB by \citet{zhang22} and within the 7$\sigma$ contour of \hess.
Considering an association with a $\sim+$55 \kms molecular cloud (MC) complex \citep{paron10},
\citet{cui21} ascribed the GeV-TeV \gray\ emission to hadronic interaction between the molecular gas by the CR protons escaped from the SNR, which was suggested to be at a distance of $\sim$3.6 kpc \citep{zhu13}.
On the contrary, \citet{zhang22} argued that the SNR component \snr\ and the HIIR component \seast\ are mutually irrelevant objects at a far distance of $\sim$10.5 kpc and a near distance of $\sim$3.4 kpc, 
respectively, and the \grays\ are likely to arise from a high-energy source hidden in the HIIR.
Additionally, the spatial associations and the well-connected GeV-TeV \gray\ spectrum which has no obvious cutoff seem to indicate a potential PeV proton accelerator in HIIR \seast.
As listed in the first LHAASO catalogue, an extended UHE \gray\ source, \lhaaso, was detected to be associated with \hess\ with a significance over 10$\sigma$ above 100\,TeV \citep{lh_cat23}.
Recently, HAWC Collaboration reported the preliminary update to their UHE catalogue, in which a point-like UHE \gray\ source, xHWC J1858$+$020, is also detected within the 6$\sigma$ contour of \hess\ \citep{hawc23}.
However, they did not provide either the spectrum or flux points.

In this work, we present a detailed LHAASO observational data analysis towards the region of the SNR-HIIR complex \snr.
Data used in this work and the corresponding results are given in Section \ref{sec:lhaaso}. 
The results of the multi-wavelength studies are shown in Section \ref{sec:multiwave} and the origin of the \gray\ emissions are discussed in Section \ref{sec:discussion}. 

\section{LHAASO data analysis and results} \label{sec:lhaaso}

LHAASO is a hybrid, large area, wide field-of-view extensive air shower array for the study of CRs and \grays\ in the sub-TeV to PeV energy range.
It consists of three detector arrays:  the 78000-$\rm{m^{2}}$ Water Cherenkov Detector Array (WCDA), the 1.3-$\rm{km^{2}}$ Kilometer Square Array (KM2A), and the Wide Field-of-view Cherenkov Telescope Array.
WCDA is sensitive to \grays\ at energies between 100\,GeV and 30\,TeV \citep{chap1_22cpc}, and KM2A operates from $\sim$20\,TeV to a few PeV \citep{cpc21k}.
LHAASO started partial operation in April 2019 and has been in full scientific operation since July 2021.
More details about the experiment can be found in \citet{cpc21k, cpc21w}.

%--------------------------------------------------------------------
\begin{table}
    \centering
    \caption{LHAASO observation data.}
    \begin{tabular}{ccc}
    \hline
        Array & Time Period (yyyy.mm.dd) & Livetime (days) \\
        \hline
        Full WCDA & 2021.03.05--2023.07.31 & 796 \\
        1/2 KM2A & 2019.12.27--2020.11.30 & 289.5 \\
        3/4 KM2A & 2020.12.01--2021.07.19 & 215.8 \\
        Full KM2A & 2021.07.20--2023.07.31 & 710.7 \\
        \hline
    \end{tabular}
    \label{tab:lh_data}
\end{table}
%--------------------------------------------------------------------

The LHAASO data used in this analysis are collected with WCDA and KM2A (see Table \ref{tab:lh_data}).
According to the number of hits ($N_{\rm hit}$), events recorded by WCDA are divided into five bins: 100--200, 200--300, 300--500, 500--800, and 800--2000, roughly corresponding to an energy range of 1 to 25\,TeV. The $\mathcal{P}_{\rm{incness}}$ parameter improved from the original definition \citep{Abeysekara_2017} is used to exclude the CR background. Events with $\mathcal{P}_{\rm{incness}}<1.1$ are used for this analysis.
Events recorded by KM2A are divided into equally logarithmic energy bins with bin width $\Delta \log_{10}E_\gamma = 0.2$ and range from 25\,TeV to several PeV. We adopt the pipeline of KM2A data analysis presented in \citet{cpc21k} with the same event selection conditions.

The sky in the region of interest (ROI) is binned into cells with a size of 0.1$^{\circ}$ and filled with detected events according to their reconstructed arrival directions for each energy bin.
The `direct integration method' \citep{fley04} is adopted to estimate the number of the CR background events.
As for the modelling of the Galactic diffuse emission (GDE), we use the same templates as that in \citet{lh_cat23}.
Dust column density measured by the PLANCK satellite \citep{2014A&A...571A..11P, 2016A&A...596A.109P} is taken as the intensity template of GDE.

The significance map in the ROI is estimated using a test statistic (TS) variable as
${\rm TS} = 2\ln({\mathcal L}/{\mathcal L}_0)$, where ${\mathcal L}$
is the maximum likelihood of a point source signal plus background hypothesis and ${\mathcal L}_0$ is the maximum likelihood of the background only hypothesis.
Assuming a power-law (PL) spectrum, the test source has an index of 2.6 in the energy range of 1--25\,TeV and 3.0 at energies above 25\,TeV.
The detection significance usually can be simplified as the square root of the TS \citep{wilks38}.
To describe the \gray\ emission observed above 1\,TeV, we fit both the WCDA and the KM2A data with a three-dimensional binned maximum likelihood method, which can determine the morphological and spectral parameters of multiple sources simultaneously.
Since the GDE is position-dependent due to the variation of column density and CR density across the Galaxy, we only fix the GDE morphology to be the same as that in \citet{lh_cat23} but set the normalization factor and spectral index of the GDE component as free parameters during the fitting process.

\subsection{WCDA analysis and results} \label{sec:wcda}

\begin{figure*}
\center
\includegraphics[angle=0,width=1.0\textwidth] {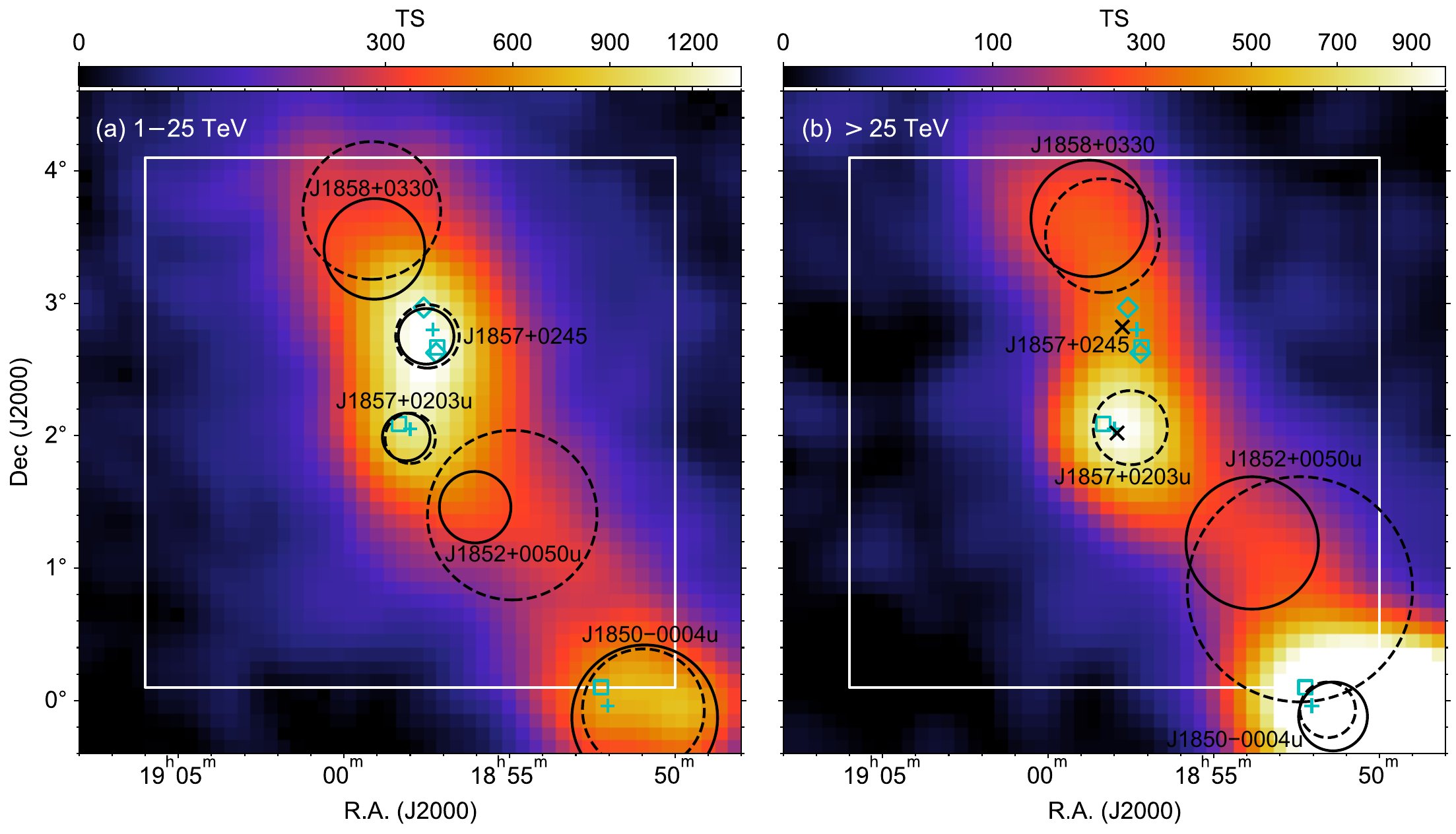}
\caption{LHAASO TS maps of a $5 \degree \times 5\degree$ region around \lhaaso. 
The white box indicates the ROI.
The black dashed circles show the WCDA or KM2A components resolved in the first LHAASO catalogue.
The cyan markers show the positions of potential TeV or multi-TeV counterparts detected by H.E.S.S. (box), MAGIC (diamond), and HAWC (cross).
(a) WCDA TS map ($1<E_{\gamma}<25$\,TeV).
The black solid circles (with radii of $r_{\rm 39}$ extensions) show the WCDA components resolved in this work.
%The colour scale is in units of significance.
(b) KM2A TS map ($E_{\gamma}>25$\,TeV).
The black crosses (corresponding to point-like sources) and black solid circles (with radii of $r_{\rm 39}$ extensions) show the KM2A components resolved in this work.
}
\label{fig:tsmaps}
\end{figure*}

For the WCDA data, we analyse a $4\degree\times4\degree$ square region centred at R.A. = 284.5$\degree$, Dec = 2.1$\degree$ in the celestial coordinate system.
Figure \ref{fig:tsmaps}(a) shows the background TS map in the ROI.
The WCDA components of five extended 1LHAASO catalogue sources (1LHAASO J1858$+$0330, 1LHAASO J1857$+$0203u, 1LHAASO J1857$+$0245, 1LHAASO J1852$+$0050u, and 1LHAASO J1850$-$0004u) may contribute to the \gray\ emission within the ROI.
Their positions and extensions resolved in the catalogue are displayed in Figure \ref{fig:tsmaps}(a) with black dashed circles. 
The extension size of the source is described by a 2D-Gaussian $r_{\rm 39}$, corresponding to 39\% of the measured events.
They all have 2D-Gaussian morphologies and PL-type spectra.
We also display the positions of the TeV or multi-TeV sources detected by H.E.S.S., MAGIC, and HAWC in Figure \ref{fig:tsmaps}.
We note that 1LHAASO J1857$+$0203u is spatially coincident with \hess\ and thus consider it our target source.
We fit the positions, extensions, and spectral parameters of all five LHAASO sources simultaneously.
The spectral parameters of the GDE are also set free.
Table \ref{tab:srcs} summarises the optimised spatial parameters for the five LHAASO sources, shown as black solid circles in Figure \ref{fig:tsmaps}(a).

%--------------------------------------------------------------------
%--------------------------------------------------------------------
%--------------------------------------------------------------------
\begin{table*}
  \centering
  \caption{Optimised parameters of the WCDA and the KM2A components in the source model.}
    \begin{tabular}{cccccc}
    \hline\hline
    1LHAASO Sources $^a$& Component& R.A.& Dec& Extension $^b$& Potential TeV/GeV Counterparts \\
 & & ($\degree$)& ($\degree$)& ($\degree$)&\\
    \hline
    1LHAASO J1858$+$0330 & WCDA& 284.77 & 3.41 & 0.38 & 4FGL J1858.0$+$0354, \\
 & KM2A& 284.69& 3.64& 0.44&4FGL J1900.4$+$0339\\
    \hline
    1LHAASO J1857$+$0245 & WCDA
& 284.38 & 2.75& 0.21 & MAGIC J1857.6$+$0297, MAGIC J1857.2$+$0263, \\
 & KM2A& 284.44& 2.82& -- &HESS J1857$+$026, 3HAWC J1857$+$027, \\
 & & & & &4FGL J1857.7$+$0246e\\
    \hline
    1LHAASO J1857$+$0203u & WCDA
& 284.53 & 1.99 & 0.18 & HESS J1858$+$020, xHWC J1858+020,\\
 & KM2A& 284.48& 2.02& -- &4FGL J1858.3$+$0209 (SrcB)\\
 & & & & & 4FGL J1857.6$+$0212 (SrcA)\\
    \hline
    1LHAASO J1852$+$0050u & WCDA
& 284.01  & 1.46 & 0.27 & 4FGL J1855.9$+$0121e,\\
 & KM2A& 283.46 & 1.19 & 0.50 &4FGL J1852.4$+$0037e\\
    \hline
    1LHAASO J1850$-$0004u & WCDA
& 282.73 & $-$0.13 & 0.55 & HESS J1852$-$000, xHWC J1852$+$000,\\
 & KM2A & 282.85& $-$0.12 & 0.26 &4FGL J1851.8$-$0007c\\
     \hline
    \end{tabular}%
    \begin{tablenotes}
        \footnotesize
        \item \textbf{Notes.}
        \item $^a$ The source names are consistent with those in the 1LHAASO catalogue.
        \item $^b$ The respective 39$\%$-containment radii.
    \end{tablenotes}
  \label{tab:srcs}%
\end{table*}%
%--------------------------------------------------------------------

We estimate the statistical position uncertainty, $\sigma_{\rm p,95,stat}$, at the 95\% confidence level using the same method as that in \citet{lh_cat23}.
Considering the pointing error only, the systematic positional uncertainty is estimated to be 0.04$\degree$ \citep{lh_cat23}.
To test the extended nature of the target source, we calculate the TS value in terms of the likelihood ratio between the best-fit extended model ($\mathcal{L}_{\rm ext}$) and the point source hypothesis ($\mathcal{L}_{\rm ps}$), defined as ${\rm TS}_{\rm ext} = 2\ln (\mathcal{L}_{\rm ext}/\mathcal{L}_{\rm ps})$.
When determining the extension of the source, a systematic bias caused by the point spread function (PSF) measurement is conservatively estimated to be $\sim0.05\degree$ \citep{lh_cat23}.
Since the ${\rm TS}_{\rm ext}$ obtained is 28.4 ($5.3\sigma$), we consider 1LHAASO J1857$+$0203u an extended source in 1--25\,TeV.
It has a statistical significance of 15.2$\sigma$ and is centred at
R.A. = 284.53$^{\circ}$ $\pm$ $0.08^{\circ}_{\rm stat}$ $\pm$ $0.04^{\circ}_{\rm sys}$,
Dec = 1.99$^{\circ}$ $\pm$ $0.09^{\circ}_{\rm stat}$ $\pm$ $0.04^{\circ}_{\rm sys}$
with an extension of $r_{\rm 39}$ = 0.18$^{\circ}$ $\pm$ $0.03^{\circ}_{\rm stat}$ $\pm$ $0.05^{\circ}_{\rm sys}$.
These results are consistent with those reported in the 1LHAASO catalogue.
Especially, the $r_{\rm 39}$ extension circle covers the \gray\ excess measured by MAGIC in this region (see \citealt{magic_J1857_2014}, Fig. 2), as well as the H.E.S.S. and HAWC sources.

\begin{figure}
\center
\includegraphics[angle=0,width=\columnwidth]{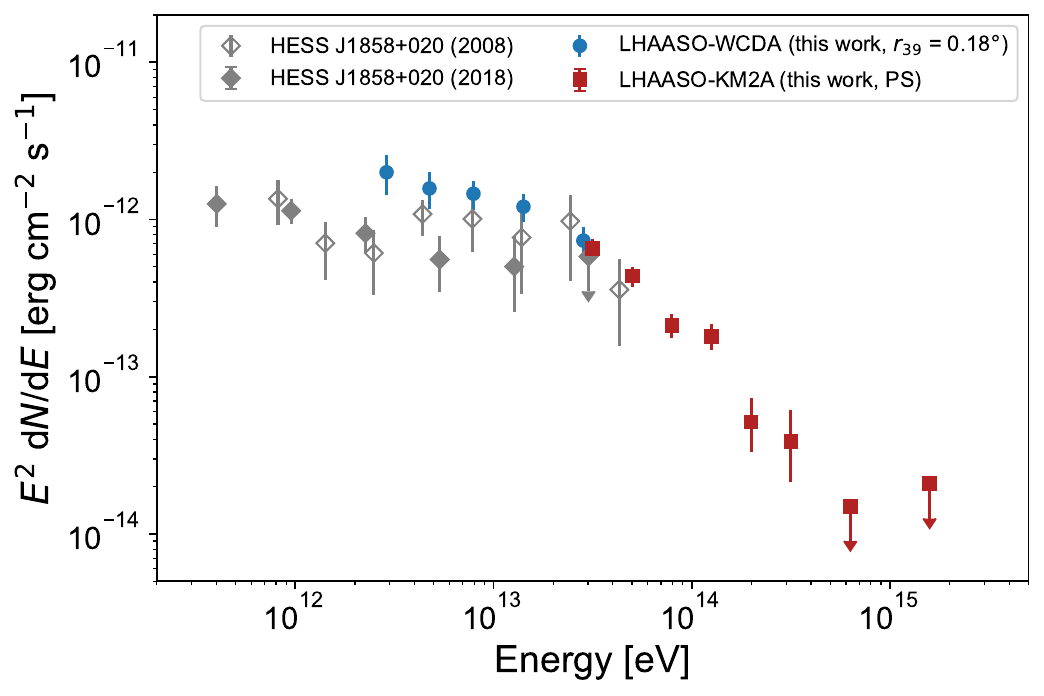}
\caption{Differential energy spectrum of the TeV source \lhaaso.
The blue dots and red squares represent the WCDA and KM2A data measured in this work, respectively. 
The combinations of statistics and systematic errors are shown.
The 95\% confidence upper limits are displayed with downward arrows.
The grey diamonds show the H.E.S.S.\ flux data points \citep{hess08, hess18}.
}
\label{fig:sed}
\end{figure}

The spectrum of the WCDA component of \lhaaso\ shows no curvature and is well fitted with a single PL: ${\rm d}N/{\rm d}E_{\gamma} = N_0 \times (E_{\gamma}/E_0)^{-\Gamma}$,
with an index of $\Gamma = 2.46 \pm 0.09_{\rm stat}$
and a differential flux normalization parameter 
$N_0 = (1.53\pm0.36_{\rm stat})\times 10^{-13}\,{\rm cm}^{-2}\,{\rm s}^{-1}\,{\rm TeV}^{-1}$ at $E_0=3$ TeV.
The derived photon index is consistent with the previous estimate by H.E.S.S.,
$\Gamma = 2.2 \pm 0.1_{\rm stat} \pm 0.2_{\rm sys}$ \citep{hess08} or
$\Gamma = 2.39 \pm 0.12_{\rm stat}$ \citep{hess18}.
We extract the spectral energy distribution (SED) in the five $N_{\rm hit}$ bins.
In each bin, we only fit the normalization parameters in the source model.
As in \citet{lh_cat23}, the systematic uncertainty is estimated to be $\sim 8\%$ on the flux for the WCDA SED measurement.
We also estimate the uncertainty caused by the imperfection model of the GDE with its spectral parameters fixed to the same values as those in the catalogue \citep{lh_cat23}.
The result shows an effect of less than 14\% on the flux.
The flux data points are shown in Figure \ref{fig:sed}.
Since the extension measured by WCDA differs from the integration radius used in the H.E.S.S. spectrum measurement (0.15$\degree$),
the flux points we obtained are slightly higher than those in \citet{hess08} but are consistent within the uncertainty range.
The difference between the measurements of WCDA and \citet{hess18} may be ascribed to the peculiarity of the background subtraction in the H.E.S.S. Galactic plane survey analysis.
\hess\ is located in a large exclusion region, the background could not be well constrained, which consequently resulted in the spectrum measurement that was not well constrained \citep{hess18}.
In our SED modelling (see Section~\ref{sec:discussion}), only the H.E.S.S. data points published in 2008 are included.

%--------------------------------------------------------------------
%--------------------------------------------------------------------

\subsection{KM2A analysis and results} \label{sec:km2a}

For the KM2A data, the same ROI as that used for WCDA is selected.
Figure \ref{fig:tsmaps}(b) shows the background TS map in the ROI.
Of the five 1LHAASO catalogue sources mentioned in Section \ref{sec:wcda}, all but 1LHAASO J1857$+$0245 have KM2A components.
Their positions and extensions resolved in the catalogue are displayed in Figure \ref{fig:tsmaps}(b) with black dashed circles.
As a null hypothesis (denoted as \textit{M0}), we add the four KM2A components and the GDE component in the source model.
As for the test models,
since 1LHAASO J1857$+$0245 is only $\sim0.7\degree$ away from the ROI centre, and its potential KM2A component may affect the fitting results, we add a fifth source as its KM2A component.
To compare with the catalogue results, we first fix the parameters of the GDE component to the same values as those in the catalogue during the fitting processes, while the parameters of the other sources are set free.
Considering the morphology (point-like/2D-Gaussian) of the target source (1LHAASO J1857$+$0203u) and the added source (1LHAASO J1857$+$0245), we test four models (\textit{M1}--\textit{M4}, see Table \ref{tab:models}), in which all the sources are assumed to have PL-type spectra.

We use the Akaike Information Criterion (AIC; \citealt{aic74,lande12}) to select the best-fit spatial model.
AIC is defined as ${\rm AIC}=2k-2\ln \mathcal{L}$, 
where $k$ is the number of free parameters in the model and $\mathcal{L}$ is the likelihood of the model tested in the analysis.
The model that minimizes the AIC is considered to be the best.
Compared with the \textit{M0} model, the $\Delta$AIC values of the models are shown in Table \ref{tab:models}.
Since the extended nature of \lhaaso\ and 1LHAASO J1857$+$0245 is not significant (${\rm TS}_{\rm ext}<9$),
we consider them to be point-like at energies above 25\,TeV.
Then, considering the GDE effects on the morphology and the flux of the sources, we free the spectral parameters of the GDE component and repeat the fit (denoted as \textit{M5--M8}).
Still, the extended nature of the two sources is not significant.
Compared to the \textit{M1} model, the \textit{M5} model shows an improvement of $3.5\sigma$ (${\rm TS} = 15.0$).
Therefore, the best-fit results of the \textit{M5} model are used in our analysis.

The KM2A component of \lhaaso\ has a statistical significance of 21.4$\sigma$ at energies above 25\,TeV and is centred at
R.A. = 284.48$^{\circ}$ $\pm$ $0.05^{\circ}_{\rm stat}$ $\pm$ $0.03^{\circ}_{\rm sys}$, Dec = 2.02$^{\circ}$ $\pm$ $0.05^{\circ}_{\rm stat}$ $\pm$ $0.03^{\circ}_{\rm sys}$.
The statistical position uncertainty $\sigma_{\rm p,95,stat}$ is estimated and the systematic positional uncertainty is estimated to be 0.03$\degree$ when the pointing error is considered \citep{lh_cat23}.
The upper limit on the extension at the 95\% confidence level is 0.18$\degree$, which is compatible with the WCDA result.
Table \ref{tab:srcs} summarises the optimised spatial parameters and the potential TeV or GeV counterparts of the five LHAASO sources.
In our analysis, the KM2A component added for 1LHAASO J1857$+$0245 has a TS value of 64 at energies above 25\,TeV, indicating a new source.
Given that, 1LHAASO J1852$+$0050u and 1LHAASO J1850$-$0004u are labelled as doubtful merging sources in the 1LHAASO catalogue, and 1LHAASO J1852$+$0050u and 1LHAASO J1858$+$0330 are affected by the GDE significantly,
it is understandable that their parameters deviate from the catalogue results.
Detailed analysis about these sources will be published separately.

\begin{figure}
\center
\includegraphics[angle=0,width=\columnwidth] {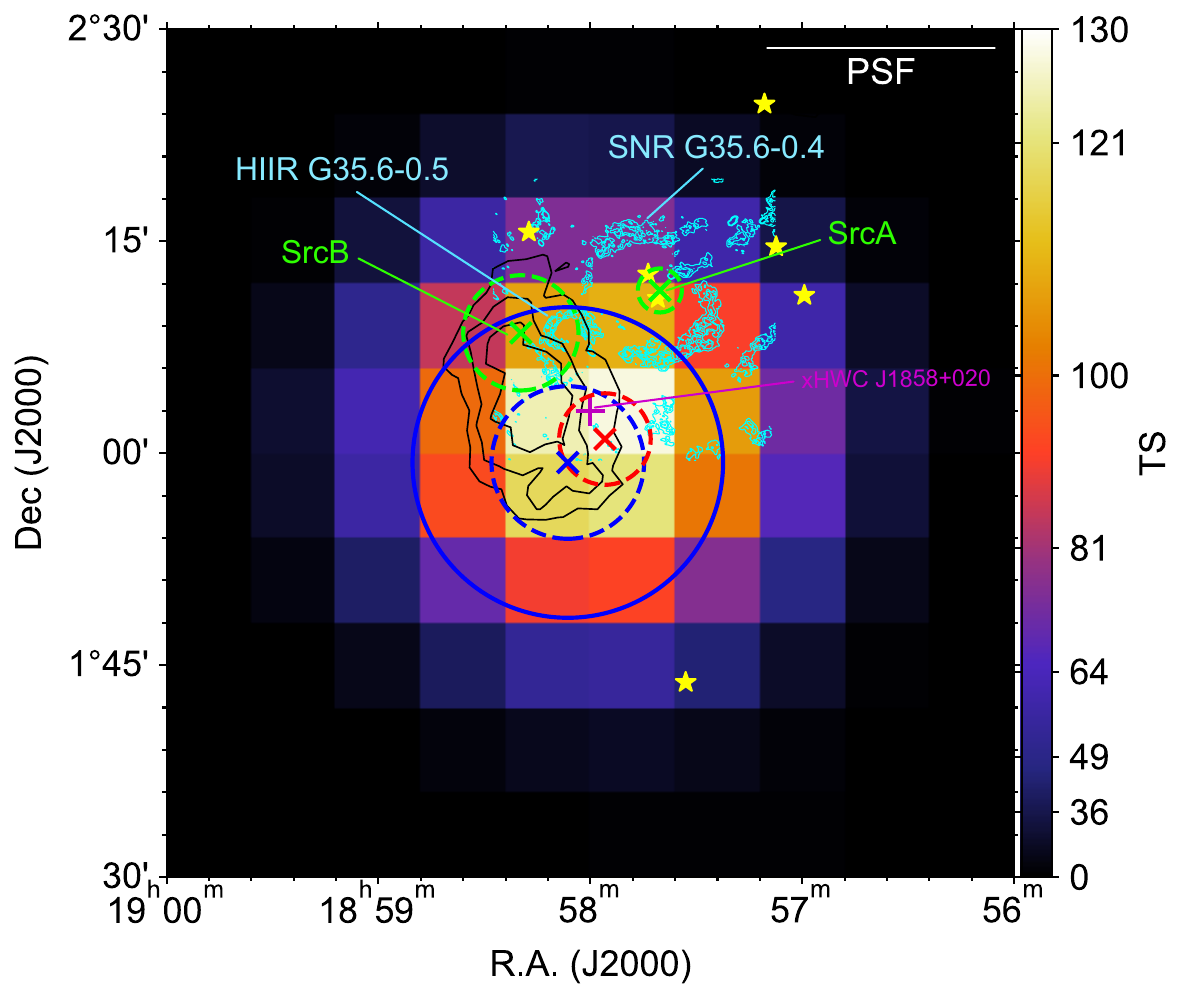}
\caption{TS map of a $1 \degree \times 1\degree$ region around \lhaaso\ as observed by LHAASO above 100\,TeV.
The blue and red crosses with $\sigma_{\rm 95}$ statistical position error circles (dashed) show the WCDA and KM2A components resolved in this work, respectively.
The solid blue circle marks the $r_{\rm 39}$ extension circle of the WCDA component.
The green crosses with $\sigma_{\rm 95}$ statistical position error circles (dashed) show the GeV \gray\ point-like sources (SrcA \& SrcB) detected by \citet{zhang22}.
The magenta cross (`+') represents the position of the UHE \gray\ point-like source detected by HAWC \citep{hawc23}.
The yellow stars represent the PSRs located within a radius of 0.5\degree\ from the TeV source.
The black contours show the TeV source \hess\ with 5$\sigma$, 6$\sigma$, and 7$\sigma$ significance \citep{hess18}.
The cyan contours indicate the GMRT 610\,MHz radio emission \citep{paredes14},
where the larger shell-like structure corresponds to SNR \snr\ and the smaller ring-shaped structure corresponds to HIIR \seast.
The line segment at the top-right corner shows the size of the PSF (68\% containment).
%The colour scale \inlet{indicates} TS values.
}
\label{fig:tsmap}
\end{figure}

Figure \ref{fig:tsmap} shows the $1^{\circ}\times1^{\circ}$ TS map around \lhaaso\ at energies above 100\,TeV, smoothed by the corresponding PSF.
The TS value for \lhaaso\ above 100\,TeV is 134, corresponding to a statistical significance of 11.6$\sigma$.
The positions of the WCDA and KM2A components of \lhaaso\ are consistent with the TeV sources xHWC J1858$+$020 and \hess\ within statistical uncertainties.
The GeV source SrcB overlaps the north of \hess\ and is possibly the GeV counterpart of the TeV sources.

The spectrum of the KM2A component of \lhaaso\ also shows no curvature and is well described by a simple PL with an index 
$\Gamma = 3.22 \pm 0.10_{\rm stat}$
and a differential flux normalization parameter 
$N_0 = (1.00\pm0.07_{\rm stat})\times 10^{-16}\,{\rm cm}^{-2}\,{\rm s}^{-1}\,{\rm TeV}^{-1}$ at $E_0=50$ TeV. 
The SED is extracted in eight logarithmically spaced energy bins.
Specifically, the width of the last two energy bins (above 400\,TeV) is $\Delta \log_{10}E_{\gamma} = 0.4$.
In each energy bin, only the normalization parameters are fitted.
If the TS value of the source in a bin is less than 4, the 95\% confidence upper limit of its flux is estimated.
As in \citet{cpc21k},
the systematic uncertainty mainly comes from the atmospheric model used in the Monte Carlo simulation.
In this case, the systematic uncertainty is estimated to be $\sim$ 7\% on the flux and 0.02 on the spectral index for the KM2A SED measurement.
The uncertainty caused by the imperfection model of the GDE is also calculated by comparing the \textit{M1} and \textit{M5} models, resulting in a deviation of less than 5\%.
Figure \ref{fig:sed} shows the differential energy spectrum of the \gray\ emission obtained in this work.
We note that the spectrum of the KM2A component can be well connected with the WCDA component and that the non-detection of extension by KM2A does not contradict the WCDA results, indicating that they are likely of the same origin.

%--------------------------------------------------------------------
%--------------------------------------------------------------------
%--------------------------------------------------------------------
\begin{table*}
  \centering
  \caption{$\Delta$AIC results for different spatial model combinations used for the KM2A analysis.}
    \begin{tabular}{cccccc}
    \hline\hline
    Models & 1LHAASO J1857$+$0203u & 1LHAASO J1857$+$0245 & GDE $^a$ & $\Delta k$ $^b$ & $\Delta$AIC $^b$ \\
    \hline
    \textit{M0} & 2D-Gaussian & --  & fixed & 0 & 0 \\
    \textit{M1} & point-like & point-like & fixed & 3 & $-$20.2 \\
    \textit{M2} & point-like & 2D-Gaussian & fixed & 4 & $-$18.8  \\
    \textit{M3} & 2D-Gaussian & point-like & fixed & 4 & $-$25.0 \\
    \textit{M4} & 2D-Gaussian & 2D-Gaussian & fixed & 5 & $-$23.4 \\
    \textit{M5} & point-like & point-like & free & 5 & $-$31.2 \\
    \textit{M6} & point-like & 2D-Gaussian & free & 6 & $-$29.4  \\
    \textit{M7} & 2D-Gaussian & point-like & free & 6 & $-$36.4 \\
    \textit{M8} & 2D-Gaussian & 2D-Gaussian & free & 7 & $-$34.4 \\
    \hline
    \end{tabular}%
    \begin{tablenotes}
        \footnotesize
        \item \textbf{Notes.}
         \item $^a$ The spectral parameters of the GDE component are fixed to the same values as those in the catalogue \citep{lh_cat23} in \textit{M0}--\textit{M4} but free in \textit{M5}--\textit{M8}.
         \item $^b$ $k$ and AIC values are provided as differences with respect to the \textit{M0} model.
    \end{tablenotes} 
  \label{tab:models}%
\end{table*}%
%--------------------------------------------------------------------

\section{Multi-wavelength studies} \label{sec:multiwave}

\subsection{Molecular environment} \label{sec:mc}

\begin{figure*}
\center
\includegraphics[angle=0,width=1.0\textwidth]{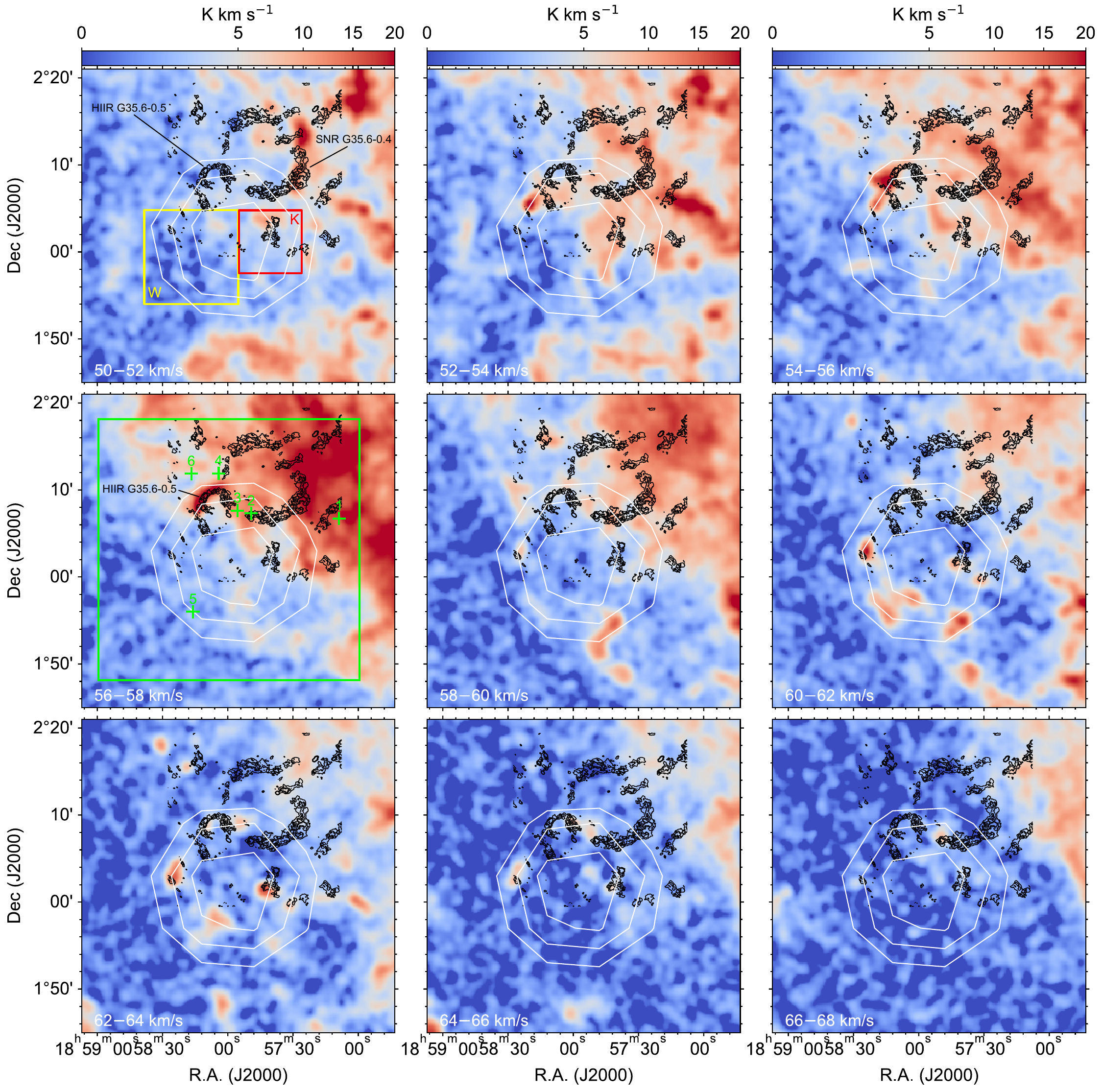}
\caption{Spatial distribution of the MCs.
FUGIN \twCO\ (\Jotz) intensity maps integrated each 2\,km\,s$^{-1}$ in the velocity range $+50$--$+68$\,km\,s$^{-1}$, overlaid with black contours of GMRT 610\,MHz radio emission (same as the cyan contours in Fig. \ref{fig:tsmap}) and white TS contours for \lhaaso\ above 100\,TeV at levels of 100, 110, and 120.
The yellow and red boxes marked with `W' and `K' in the top left panel show the regions that are used to estimate the parameters of the $+50$--$+70$ km s$^{-1}$ molecular gases.
The green box in the middle left panel
($+$56--$+$58 km s$^{-1}$, corresponds to a near distance of $3.4\pm0.4$ kpc)
marks the region where we search for OB stars, and the green crosses mark the positions of the OB star candidates.
}
\label{fig:co}
\end{figure*}

To investigate the potential molecular gases that may contribute to the hadronic emission,
we use the archival CO data of the FOREST Unbiased Galactic plane Imaging survey with the Nobeyama 45 m telescope (FUGIN; \citealt{co17}) observation.
The data has an angular resolution of 20$^\prime$$^\prime$ and an average RMS of 1.5 K at a velocity resolution of 0.65 km s$^{-1}$.
Figure \ref{fig:co} shows the spatial distribution of the \twCO\ (\Jotz) emission in velocity range $\sim +50$ to $+68$ km s$^{-1}$, with velocity intervals 2 km s$^{-1}$.
Some molecular clumps appear in morphological agreement with the detected TeV \gray\ emissions.
As shown in Figure \ref{fig:co}, we denote the clumps in two box regions as `W' and `K'.
The centre positions and sizes of the clouds are listed in Table \ref{tab:mc}.
When estimating the molecular column density and the mass of the molecular gas, local thermodynamic equilibrium for the molecular gas and optically thick conditions for the \twCO\ (\Jotz) line are assumed.
The excitation temperature is calculated with $T_{\rm ex}={5.53}/{\ln[1+{5.53}/({T_{\rm peak}(^{12}{\rm CO})+0.819})]}$ K \citep{naga98},
where $T_{\rm peak}(^{12}{\rm CO})$ is the peak temperature of the maximum \twCO\ (\Jotz) emission point in the cloud region.
The column density is calculated via 
$N({\rm H}_2) =1.49\times10^{20}\ {W(^{13}{\rm CO})}/[{1-\exp (-5.29/T_{\rm ex})}]$ cm$^{-2}$ \citep{naga98},
where $W$($^{13}$CO) is the integrated intensity of the \thCO\ (\Jotz) line in the velocity range $\sim+50$ to $\sim+70$ km s$^{-1}$.
The abundance ratio [H$_2$]/[$^{13}$CO] is taken as $7\times10^5$ \citep{frerk82}.
After estimating the $N({\rm H}_2)$ value in each pixel of the corresponding regions, the average values for the clouds are obtained.
The relation $M=2.8m_{\rm H}N({\rm H}_2)A$ is then used to derive the mass of the molecular gas,
where $m_{\rm H}$ is the mass of a hydrogen atom and $A$ is the cross-sectional area of the corresponding clouds.
For comparison, both the near ($\sim$3.4 kpc) and far ($\sim$10.5 kpc) distances \citep{zhang22} are used in the estimation.
The clouds are simply approximated as cubes when calculating the average number densities of hydrogen nuclei ($n_{\rm H}$). 
The derived parameters of the molecular clouds are summarised in Table \ref{tab:mc}.

%--------------------------------------------------------------------
%--------------------------------------------------------------------
 
%--------------------------------------------------------------------
\begin{table*}[ht]
\setlength{\tabcolsep}{8pt}
\center
  \caption{Properties of the molecular and atomic clouds.}
    \begin{tabular}{cccccccc}
    \hline\hline
    Clouds& (R.A., Dec) $^a$ & Size & $N$(H$_2$) or $N$(HI) & $d$ & $M$ & $n_{\rm H}$ & ${T_{\rm ex}}$\\
 & ($\degree$)& ($\degree$)& (10$^{21}$ cm$^{-2}$)&  (kpc)&($10^3M_{\odot}$) & (cm$^{-3}$)&(K)\\
    \hline
  %  \hline
   W & (284.57, 1.99) & 0.18 & 1.6 &  10.5&39.3& 30& 23.4\\
 & & & & 3.4& 4.1& 100 &\\
    \hline
    K & (284.42, 2.02) & 0.12 & 2.0 &  10.5&21.5& 60& 18.8\\
 & & & & 3.4& 2.3& 180 &\\
     \hline
     A & (284.53, 1.99) & 0.26 & 3.0 & 10.5 & 76.3 & 20 & --\\
 & & & & 3.4 & 8.0 & 60 &\\
 \hline
    \end{tabular}%
    \begin{tablenotes}
		\footnotesize
		\item \textbf{Notes.}
		\item $^a$ centre of the cloud.
  \end{tablenotes}
  \label{tab:mc}%
\end{table*}%
%--------------------------------------------------------------------

\subsection{Atomic clouds} \label{sec:h1}

\begin{figure}
\center
\includegraphics[angle=0,width=\columnwidth] {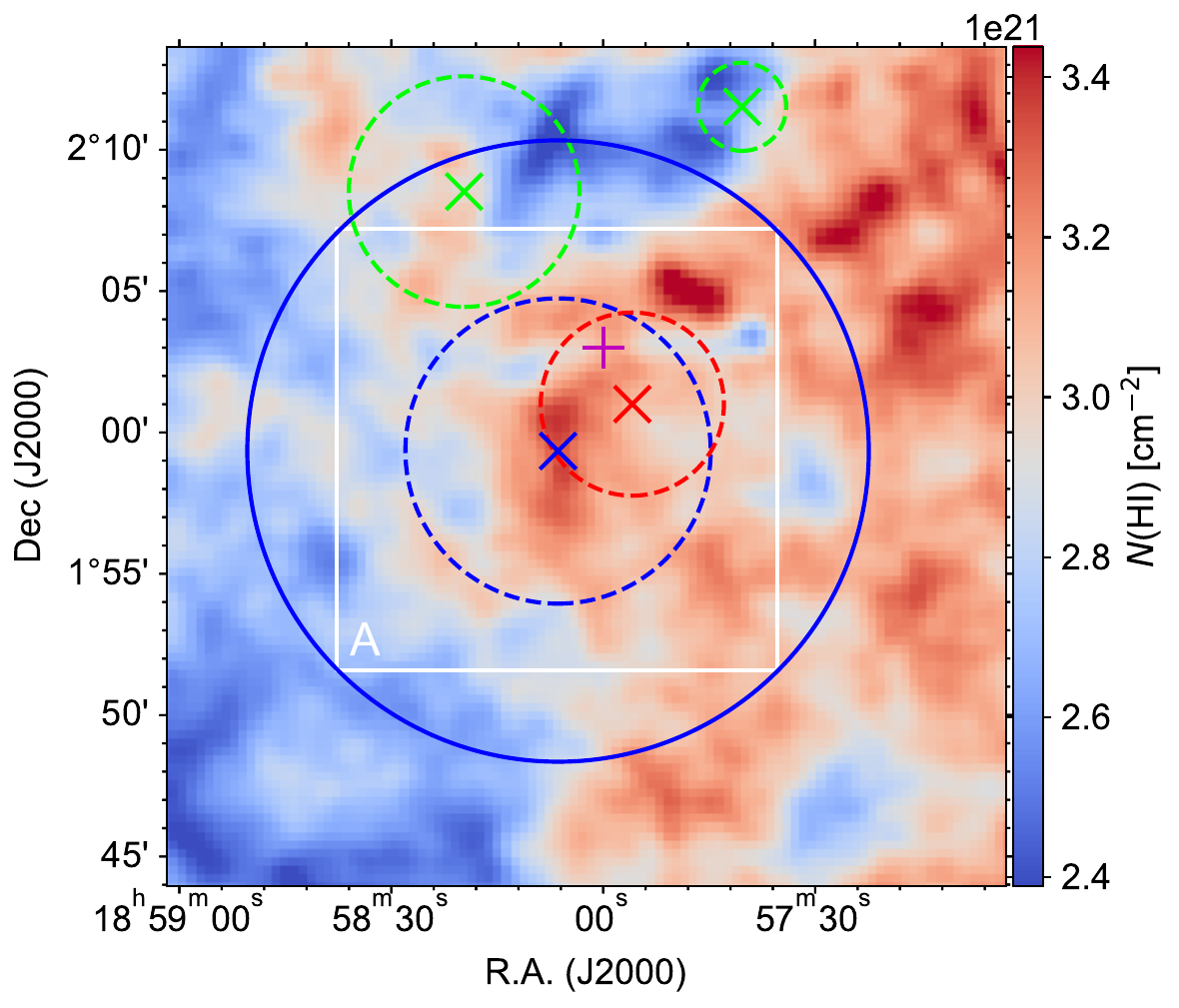}
\caption{HI column density map calculated in the velocity range of $+50$--$+70$ km s$^{-1}$.
The crosses and circles are the same as those in Figure \ref{fig:tsmap}.
The white box marked with `A' shows the region used to estimate the parameters of the $+50$--$+70$ km s$^{-1}$ atomic gases.
}
\label{fig:h1}
\end{figure}

To estimate the parameters of the atomic clouds that may be associated with the \gray\ emissions,
we obtain the 1420 MHz HI line emission data from the Very Large Array Galactic Plane Survey \citep{vla06}.
The synthesized beam for the HI spectral line images is $18^{\prime\prime}$ and the radial velocity resolution is 0.824 km s$^{-1}$.
Figure \ref{fig:h1} shows the column density map of the HI line emission
calculated in the velocity range of $+50$--$+70$ km s$^{-1}$.
The atomic hydrogen column density is measured using an HI intensity-mass conversion factor of $X_{\rm HI}=1.823 \times 10^{18}\, {\rm cm^{-2}/(K\,km\,s^{-1})}$ \citep{roh04}.
The centre gases appear to be spatially consistent with the detected TeV \gray\ emissions.
We mark the gases in a box region as `A' (see Figure \ref{fig:h1}), of which the centre position and size are listed in Table \ref{tab:mc}.
Similar to the estimation methods in Section \ref{sec:mc}, the average column density, mass, and number density of hydrogen nuclei for cloud `A' are also obtained and summarised in Table \ref{tab:mc}.

%--------------------------------------------------------------------
%--------------------------------------------------------------------

\subsection{Search for OB stars}\label{subsec:obstars}
%--------------------------------------------------------------------
\begin{table*}
\setlength{\tabcolsep}{5pt}
     \centering
     \caption{Candidate-B stars around HIIR \seast.}
     \begin{tabular}{cccccccccc}
     \hline\hline
         No.&  ID&  (R.A., Dec)&  $G$&  $A_G$&  $BP$&  $RP$&  Parallax&  Distance& $T_{\rm eff}$ \\
         {} & {} & (\degree) & (mag) & (mag) & (mag) & (mag) & (mas) & (pc) & (K) \\
         \hline
         1 & 4267138008163989376 & (284.289202, 2.109622) & 15.44 & 3.32 & 16.33 & 14.49 & 0.608$\pm$0.038 & 3383.17 & 9142.34 \\
         2 & 4268591631256410368 & (284.457488, 2.119467) & 15.36 & 3.38 & 16.26 & 14.41 & 0.481$\pm$0.035 & 3306.22 & 9267.10 \\
         3 & 4268591974853788544 & (284.483678, 2.124783) & 16.03 & 3.66 & 16.98 & 14.98 & 0.578$\pm$0.137 & 3877.79 & 9386.42 \\
         4 & 4268640731323327488 & (284.519704, 2.196094) & 15.41 & 2.91 & 16.16 & 14.56 & 0.549$\pm$0.034 & 3755.30 & 9198.58 \\
         5 & 4267084368314521088 & (284.569054, 1.931037) & 15.32 & 3.30 & 16.20 & 14.38 & 0.517$\pm$0.029 & 3514.04 & 9039.63 \\
         6 & 4268594139517304448 & (284.572168, 2.195489) & 15.34 & 2.93 & 16.08 & 14.48 & 0.550$\pm$0.034 & 3152.76 & 9130.34 \\
     \hline
     \end{tabular}
     \label{tab:gaia}
\end{table*}
%--------------------------------------------------------------------
It is suggested that particles can be accelerated by the colliding winds of young massive clusters \citep{Aharonian19} which are the ionizing sources of HIIRs.
O- and B-type (OB) stars are massive stars that have strong stellar winds. 
To find massive stars (OB stars) that may be associated with HIIR \seast,
we first search in the OB-star candidate catalogue selected from the VST Photometric H$\alpha$ Survey data release 2 (VPHAS$+$ DR2) \citep{chen19} and the available catalogue of all O-B2 stars compiled from the literature \citep{chen19} but obtain no results.
Then, we select all sources from the \textit{Gaia} DR3 database (the \texttt{gaia\_source} table, \citealt{gaia16, gaia23_all}) in a $0.5\degree \times 0.5\degree$ area, centred on ${\rm R.A.} = 284.50\degree$, ${\rm Dec} = 2.05\degree$ (see the green box in the middle left panel of Figure \ref{fig:co}).
Then we select the stars that received a \texttt{spectraltype\_esphs} tag $\in$ [`O', `B'] in the \texttt{astrophysical\_parameters} table \citep{gaia23_ap},
and fix the effective temperature $T_{\rm eff}$ threshold at 9000 K. 
After filtering, 26 sources remained in the list of candidate-OB stars.
Among these, seven sources are distributed in a distance ($d$) range between 2.8\,kpc and 4.0\,kpc, which covers the near kinematic distance of HIIR \seast\ ($3.4\pm0.4$\,kpc).
For a further constraint, we consider the sources with absolute $G$ magnitude $M_G = G - 5(\log_{10} d - 1) - A_G<0$ \citep{chen19}, where $A_G$ is the interstellar extinction.
This results in a total of six candidate-B stars (see Table \ref{tab:gaia} and Figure \ref{fig:co}), among which four are in the \texttt{gold\_sample\_oba\_stars} table \citep{gaia23_ap}.
Unfortunately, the candidates appear %sporadic
scattered, and none are near the centre of HIIR \seast.
These isolated candidate-B stars are incapable of accelerating CRs to high energies.
We also investigate the \textit{Gaia} colour-absolute magnitude diagram, as well as other observations, such as VPHAS$+$ DR3 and 2MASS, but no positive results appear.
We therefore conclude that, based on the data available to date, no OB associations or massive star clusters have been found to be associated with HIIR \seast.

\subsection{Radio and X-ray observations}\label{subsec:radx}

We explore the radio observations towards the \lhaaso\ region with the survey data:
the 1--2\,GHz HI/OH/Recombination line survey (THOR; \citealt{beuther_2016,wang_2020}) combined with the Very Large Array Galactic Plane Survey (VGPS; \citealt{vla06}), which is called the THOR$+$VGPS\footnote{\url{https://www2.mpia-hd.mpg.de/thor/Data_\%26_Publications.html}},
and the Effelsberg 11\,cm ($\sim 2.7$\,GHz) survey of the Galactic plane by \citet{reich_1984}\footnote{\url{https://www3.mpifr-bonn.mpg.de/survey.html}}.
No significant radio emissions were detected in the central region of \lhaaso\ at 1.4\,GHz and 2.7\,GHz.
We estimate the upper limits of the flux densities for a source with a radius of 0.18$\degree$ ($=r_{\rm 39}$) centred at R.A. = 284.53$\degree$, Dec = 1.99$\degree$.
To remove the contamination from HIIR \seast\ and SNR \snr\, we first calculate a circular region with a radius of 0.09$\degree$ (0.5$r_{\rm 39}$) and then scale the results.
%1$\sigma$, 0.09$\degree$: 8.2\,Jy @ 1.4\,GHz, 3.2\Jy @ 2.7\,GHz
The 2$\sigma$ upper limits of the flux densities at 1.4\,GHz and 2.7\,GHz are 65.5\,Jy and 25.3\,Jy, respectively.

\citet{paredes14} analysed the \textit{Chandra} X-ray (0.4--7\,keV) data in the \hess\ region and found no extended X-ray emission.
They obtained the 2$\sigma$ upper limit to be $\sim1\times10^{-13}$\,erg\,cm$^{-2}$\,s$^{-1}$
for a source with a $6^\prime\times6^\prime$ extension and an X-ray spectrum of a PL of photon-index $\Gamma=2$, based on the statistical fluctuation of the background of the field.
Here we scale the result to the $r_{\rm 39}$ extension of \lhaaso\ and obtain $\sim1\times10^{-12}$\,erg\,cm$^{-2}$\,s$^{-1}$.
No other X-ray telescopes have observed or detected emission in this region.

\section{Discussion} \label{sec:discussion}

In the above LHAASO data analysis, we found that 1LHAASO J1857$+$0203u is an extended source in the WCDA band (1--25\,TeV) which almost covers the TeV source HESS J1858$+$020 and the GeV source SrcB (see Figure~\ref{fig:tsmap}).
In the KM2A band ($>25$\,TeV), however, it is a point-like source and has an angular separation of $0.18^{\circ}$ to SrcB, without overlap between the position error circles of them.
%Considering the PSF of KM2A, 
Considering the position uncertainties, 
it is not easy to identify whether SrcB, HESS J1858+020, and 1LHAASO J1857+0203u have the same origin.
If this is the case, as suggested in \citet{zhang22}, HIIR \seast\ would be a likely source of the GeV/TeV--PeV \gray\ emissions.
Meanwhile, 1LHAASO J1857+0203u projectively appears close to SNR G35.6$-$0.4, with an angular distance less than $0.2^{\circ}$.
Hence, there is a possibility that 1LHAASO J1857+0203u is unrelated to SrcB but is produced by the illumination of MCs by the escaped protons from the SNR.
In the latter case, SrcB would be an independent source associated with HIIR \seast.
Furthermore, given the deviations of the central positions of the GeV and TeV \gray\ sources, it is also possible that the TeV emission is neither associated with the HIIR nor the SNR. 
An undetected PWN scenario would then be considered.
We searched the SIMBAD Astronomical Database \citep{wenger00}\footnote{\url{http://simbad.u-strasbg.fr/simbad/sim-fcoo}} and found no binary system or other plausible counterpart.
Therefore, we discuss the above three scenarios in this section.

\subsection{HII region scenario}

Given the association of the GeV $\gamma$-ray position with the MCs around the HIIR \seast\ at the near distance $\sim3.4$\,kpc, \citet{zhang22} suggested the GeV-TeV spectral points that seem to be smoothly connected could be explained by a hadronic interaction.
This notion is extended here to the combination of the \textit{Fermi}-LAT GeV data, the WCDA TeV data, and the KM2A sub-PeV data.
In this scenario, the \gray\ emissions from GeV to sub-PeV around the radio complex \snr\ are considered powered by a potential accelerator hidden in the HIIR \seast.
The WCDA component of 1LHAASO J1857$+$0203u and HESS J1858$+$020 are treated as the same \gray\ source, and the SrcB and the KM2A component are the GeV and sub-PeV counterparts, respectively.

\begin{figure}
\center
\includegraphics[angle=0,width=\columnwidth]{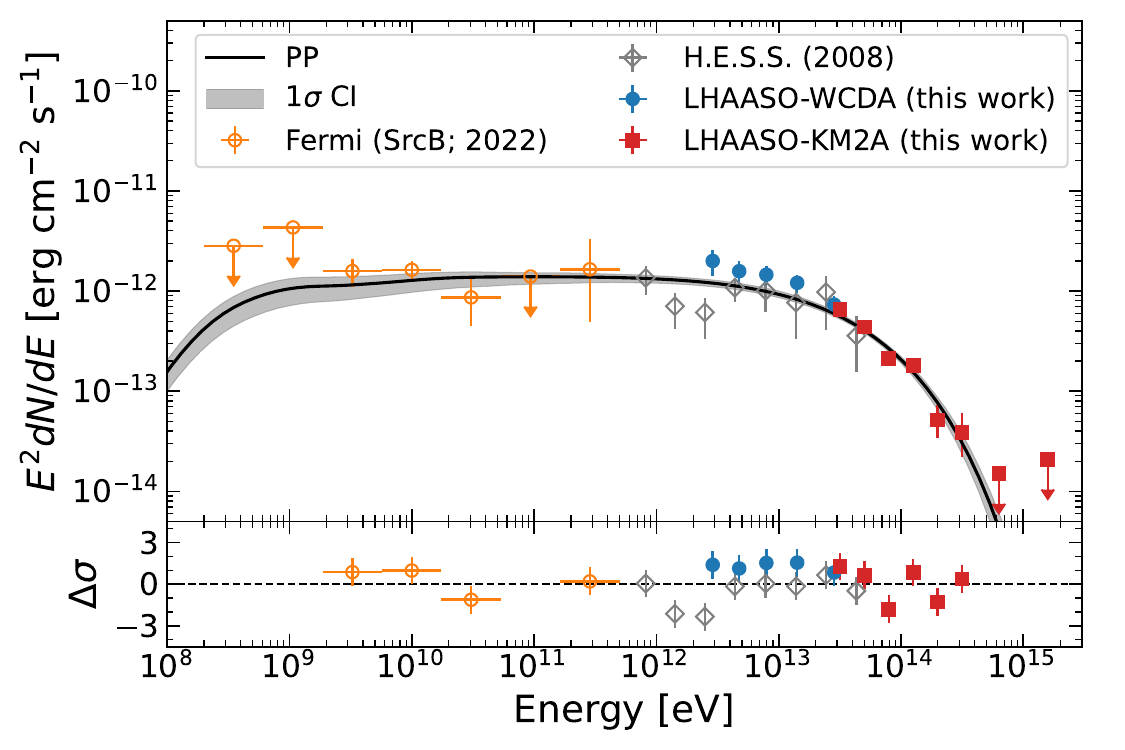}
\caption{Fitted SED for the HII region scenario. The pion-decay \grays\ with $1\sigma$ confidence interval is represented by the black line. The Fermi data (orange circle) for SrcB are adopted from \citet{zhang22}.}
\label{fig:sed_hii}
\end{figure}

We assume the accelerated protons have a PL distribution in energy with an exponential cutoff $E_{\rm p,cut}$:
${\rm d}N_{\rm p}/{\rm d}E_{\rm p} \propto E_{\rm p}^{-\alpha_{\rm p}} {\rm exp}(-E_{\rm p}/E_{\rm p,cut})$,
where $E_{\rm p}$ is the particle energy, and $\alpha_{\rm p}$ is the PL index.
The normalization is determined by the total energy in particles with energies above 1\,GeV, $W_{\rm p}$.
We use the \texttt{Naima} package (version 0.9.1, \citealt{zab15}) to fit the GeV--PeV energy spectrum.
In the fitting process, the $\gamma-\gamma$ absorption is not considered due to the negligible influence on the flux for the distance of 3.4 kpc (see Appendix \ref{sec:absorption}).
By adopting an average atomic hydrogen number density $n_{\rm t}\sim 600\ {\rm cm}^{-3}$ of the target gas for the p-p interaction \citep{zhang22},
we obtain 
$\alpha_{\rm p}=2.04^{+0.04}_{-0.05}$, $E_{\rm p,cut}=445^{+83}_{-67}$\,TeV, 
and $W_{\rm p} = 1.34^{+0.27}_{-0.25}\times 10^{47}(n_{\rm t}/600\ {\rm cm^{-3}})^{-1}$\,erg.
The fitted SED is shown in Figure~\ref{fig:sed_hii}.

Although no massive stars have yet been detected in HIIR \seast\ (see Section \ref{subsec:obstars}),
it cannot be ruled out that they are embedded in the MCs.
Here we assume that the energetic protons are accelerated by the stellar winds of the massive stars and discuss the possibility from the energy budget side.
The proton luminosity $L_{\rm p}(>1\ {\rm GeV})$ can be roughly derived from $\int_{\rm 1 GeV}^{+\infty} E_{\rm p} {\rm d}N_{\rm p}/{\rm d}E_{\rm p}/{\rm min}(\tau_{\rm pp},t_{\rm c}) {\rm d}E_{\rm p}$, where $\tau_{\rm pp}$ and $t_{\rm c}$ are the proton's lifetime and confinement time, respectively. 
For the hadronic process, the lifetime is $\tau_{\rm pp}=10^{5} (n_{\rm t} /600\ {\rm cm^{-3}})^{-1}$yr.
The confinement time can be estimated by $t_{\rm c}=L^2/6D$ \citep{aharonian96}, where $L$ and $D$ are the source physical size 
($\sim$0.2$^{\circ}$) 
and the diffusion coefficient, respectively.
We assume that the diffusion coefficient has a general form $D=\chi 10^{28}(\Ep/10\ {\rm GeV})^{\delta}\ {\rm cm^2\ s^{-1}}$, where both $\chi$ and $\delta$ are dimensionless constant.
The particular values of $\delta = 1/3$ and 1 correspond to the Kolmogorov- and Bohm-like diffusion regimes, respectively.
For the CR propagation models, $\chi=1$ and $\delta=0.3$--0.6 are required \citep[e.g.][]{Berezinskii1990}.
While recent observations suggest that the diffusion coefficient around the particle acceleration site may be significantly lower than the average value in Galaxy, giving $\chi\le0.01$ \citep[e.g.][]{fujita10,ohira11, li12c,abeysekara17}.
Considering this fact, the injection luminosity of protons is 
$L_{\rm p} \sim (4$--$50)\times10^{35}(\chi/0.01)(L/10\ {\rm pc})^{-2}\ {\rm erg\ s^{-1}}$ for $\delta =0.3-0.6$.
For a typical OB-type star, it has a strong stellar wind with a mass loss rate of order $10^{-6}\ M_{\odot}\ {\rm yr}^{-1}$ and a velocity of order $10^3\ {\rm km\ s^{-1}}$,
giving total mechanical energy in wind $\sim 10^{35}\ {\rm erg\ s^{-1}}$.
Assuming that a fraction $\eta_{\rm w}\sim10\%$ of the mechanical energy is used to accelerate protons, the required number of massive stars is $N\sim40$--$500(\eta_{\rm w}/0.1)^{-1}(\chi/0.01)(L/10\ \rm{pc})^{-2}$.
This star number within a region of $\sim 10$~pc can be satisfied in the young massive star clusters or associations with a typical radius of a few pc and $10^3$--$10^6$ stars \citep{Zwart2010}.
For instance, there are $\sim 300$ stars with a mass above 15 $M_\odot$ for a young massive star cluster with a total mass of $10^5\ M_\odot$ and a Salpeter-like mass function \citep{Salpeter1955} in the mass range from 0.1 to 100 $M_{\odot}.$
Thus deep searches for OB stars in this region are still needed.
If there are enough massive stars and/or strong suppression of the diffusion coefficient, the stellar winds could supply the required energy to accelerate particles.

A number of protostars, with at least one among them possibly of high mass nature, were found in HIIR \seast\ \citep{paron10, paron11}.
Although no signature of outflows was confirmed \citep{paron11},
the apparent elongation of the region may indicate a powerful outflow, which could be masked in radio by strong free-free absorption \citep{paredes14}. 
In addition, the elongated CO structure and several X-ray sources with positions roughly aligned with the former indicate that there is an active star-formation region \citep{paredes14}.
The outflow and the entire star formation region could also be the origin of, or contribute to, the \gray\ emissions observed.

\subsection{Clouds illuminated by particles escaped from SNR}

\begin{figure}
\center
\includegraphics[angle=0,width=\columnwidth]{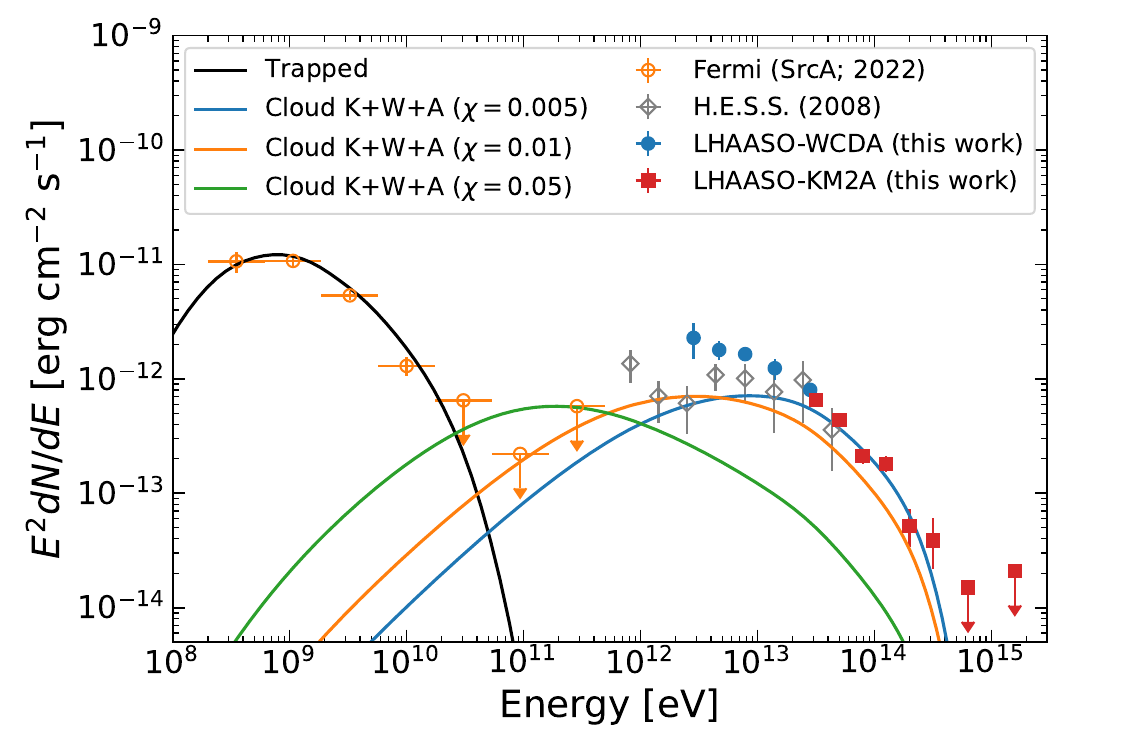}
\caption{SED of the SNR scenario including the $\gamma-\gamma$ absorption. The black solid line represents the \grays\ from the trapped protons, which explains the GeV data of SrcA \citep{zhang22}. The \gray\ emission produced by the escaped protons is displayed in blue, orange, and green solid lines for $\chi=0.005$, 0.01, and 0.05, respectively.
}
\label{fig:sed_snr}
\end{figure}

The second case is that the TeV--PeV \grays\ arise from the MCs around SNR \snr\ illuminated by the diffusing protons that escaped from the SNR shock wave. 
The estimated far kinematic distance of Clouds K, W, and A with $v_{\rm LSR}\sim60\pm10\ {\rm km\ s^{-1}}$ is $\sim$10~kpc \citep{wenger18,reid14}, which is very close to that of SNR \snr\ ($\sim$ 10.5 kpc, \citealt{zhang22}).
Thus, there is a possibility that the three clouds are at the same distance as the SNR and are located in the vicinity of the SNR. 
While some of the escaped protons illuminate the nearby non-contact clouds,
the trapped protons at the current stage also can generate \gray\ emission via bombarding the interacting MCs, which may explain the GeV source SrcA in \citet{zhang22}.
We calculate the expected \gray\ fluxes according to the so-called isotropic delta-escape model described in \citet{gabici2009} and compare them with the observational data.
In the calculation, the $\gamma-\gamma$ absorption is included due to the large distance (see Appendix \ref{sec:absorption}).

The distribution function of escaped protons 
without energy losses
at any given distance $R$ from the SNR centre and at any given time $t$ is \citep{atoyan95,gabici2009}
\begin{equation}
f_{\rm esc}(\Ep, R, t)=\frac{A_0 \Ep^{-\alpha_{\rm p}}}{\pi^{3/2}R_{\rm d}^{3/2}}{\rm exp}\left(  -\frac{R^2}{R_{\rm d}^2} \right),
\end{equation}
where $\alpha_{\rm p}$ and $\Ep$ are the PL index and the proton energy, respectively.
The normalization factor, $A_0$, is derived from $\eta E_{\rm SN}=\int^{E_{\rm p,max}}_{\rm 1\ GeV}\Ep A_0 \Ep^{-\alpha_{\rm p}}d\Ep$, where $E_{\rm SN}$ is the explosion energy, $\eta$ is the the energy conversion efficiency, and $E_{\rm p,max}$ is the maximum energy of the accelerated protons.
The diffusion length is 
\begin{equation}
    R_{\rm d}=\sqrt{4D(\Ep)(t-t_{\rm esc}(\Ep))},
\end{equation}
where $t_{\rm esc}(\Ep)=t_{\rm sed}(\Ep/E_{\rm p,max})^{-1/\mu}$ is the parameterised escape time when the protons of energy $\Ep$ escape from the SNR \citep{gabici2009} with $t_{\rm sed}$ the beginning time of the Sedov phase and $\mu$ a free parameter. 
In the above equation, $D(\Ep)$ is the energy-dependent diffusion coefficient and the form in the CR propagation models $D(\Ep)=\chi 10^{28}(\Ep/10\ {\rm GeV})^{\delta}$ with $\delta=0.3$--0.6 \citep{Berezinskii1990} is adopted here.
The distribution of the protons confined in the SNR is
\begin{equation}
    dN_{\rm p}/d\Ep=A_0 \Ep^{-\alpha_{\rm p}}{\rm exp}(-\Ep/E_{\rm p,trap}),
\end{equation}
where $E_{\rm p,trap}=E_{\rm p,max}(t/t_{\rm sed})^{-\mu}$.

Given the distribution of the escaped protons, the flux of the \gray\ emission from a cloud is 
% \citep[e.g.,][]{aharonian96}
\begin{equation}
    \frac{dN}{dE_{\gamma}}=\frac{M_{\rm c}}{m_{\rm H}}\frac{\epsilon_{\gamma}}{4\pi d^2},
\end{equation}
where $M_{\rm c}$ and $m_{\rm H}$ are the mass of the cloud and the mass of the hydrogen atom, respectively, $\epsilon_{\gamma}$ is the \gray\ emissivity per hydrogen atom and is calculated based on the method developed by \citet{kafexhiu2014}.

SNR \snr\ almost has a circular shape in the sky plane with an average radius of 15\,pc at a distance of 10.5\,kpc.
Although there are the associated MCs at the western boundary suggested by \citet{zhang22}, the average ambient (or inter-cloud) density $n_0\sim1\ {\rm cm}^{-3}$ is assumed to estimate the dynamical age.
Assuming an explosion energy $E_{\rm SN}=10^{51}$\,erg and the ejecta mass $M_{\rm ej}=2\ M_{\odot}$, the beginning time of the Sedov phase is $t_{\rm sed}\sim 330$\,yr and the SNR age can be estimated as $t_{\rm SNR}\sim16$\,kyr based on the Sedov solution.
The separation between Cloud K/W/A and the SNR centre is 9.6$^\prime$/13.9$^\prime$/13.1$^\prime$, corresponding to a projective distance of $\sim$29/43/40\,pc, respectively.
The true distance of clouds from the SNR centre should not be smaller than the projective distance.
In our calculation, $R_{\rm K}=30$\,pc, $R_{\rm W}=45$\,pc, and $R_{\rm A}=45$\,pc are adopted for the Cloud K, W, and A, respectively.

For the accelerated protons, we assume the index $\alpha_{\rm p}=2$, $\eta=10\%$, and $E_{\rm p,max}=1$\,PeV.
Then we apply the above escape model to the observational data.
We first fit the GeV data of SrcA (see the black line in Figure~\ref{fig:sed_snr}), resulting in $\mu\approx2.68$ and $n_{\rm t}=200\ {\rm cm}^{-3}$.
The fitted target density is consistent with the SNR-MC interaction scenario \citep{zhang22}.
For $\mu\approx2.68$, corresponding to $E_{\rm p,trap}\sim30$\,GeV at the current SNR age, it means that the protons with energy below 30\,GeV are still confined in the SNR region.

Then, the \gray\ emission from the nearby MCs illuminated by the escaped protons can be calculated, which is determined by two parameters $\delta$ and $\chi$.
In Figure~\ref{fig:sed_snr}, as an example, we display the expected \gray\ flux from Clouds K, W, and A with $\chi=0.005$ (blue), 0.01 (orange), and 0.05 (green) for the Kolgomorov-type diffusion $\delta=1/3$.
The peak energy of SED can shift down to the TeV energies for larger $\chi$, while the flux around $\sim$100\,TeV is suppressed.
This is because the protons that produce $\sim$100\,TeV \grays\ have effectively passed through the cloud due to the larger diffusion coefficient \citep{aharonian96}.
In addition, $\delta$ and $\chi$ are degenerated, resulting in the same situation for other values of $\delta$.
As can be seen, the predicted \gray\ emission from the escaped protons can not explain the LHAASO data unless
the nearby clouds have a larger mass (e.g. the mass of Cloud K is amplified by 4 times, see Appendix \ref{sec:snr_ap}).
Therefore, the isotropic escape model can be ruled out due to the low total masses and/or the large separation from SNR of Clouds K, W, and A.

If the diffusion process is not isotropic, however, the required cloud mass can be smaller.
For the bipolar diffusion in which the escaped protons diffuse from the SNR only within two cones and one happens to cover Clouds K, W, and A, the energy density of the protons can be roughly amplified by a factor of $10.7(\Omega/0.59)^{-1}$, where $\Omega$ is the solid angle of a cone and its apex angle $50^{\circ}$ is adopted here.

\subsection{PWN scenario}

\begin{figure}
\center
\includegraphics[angle=0,width=\columnwidth]{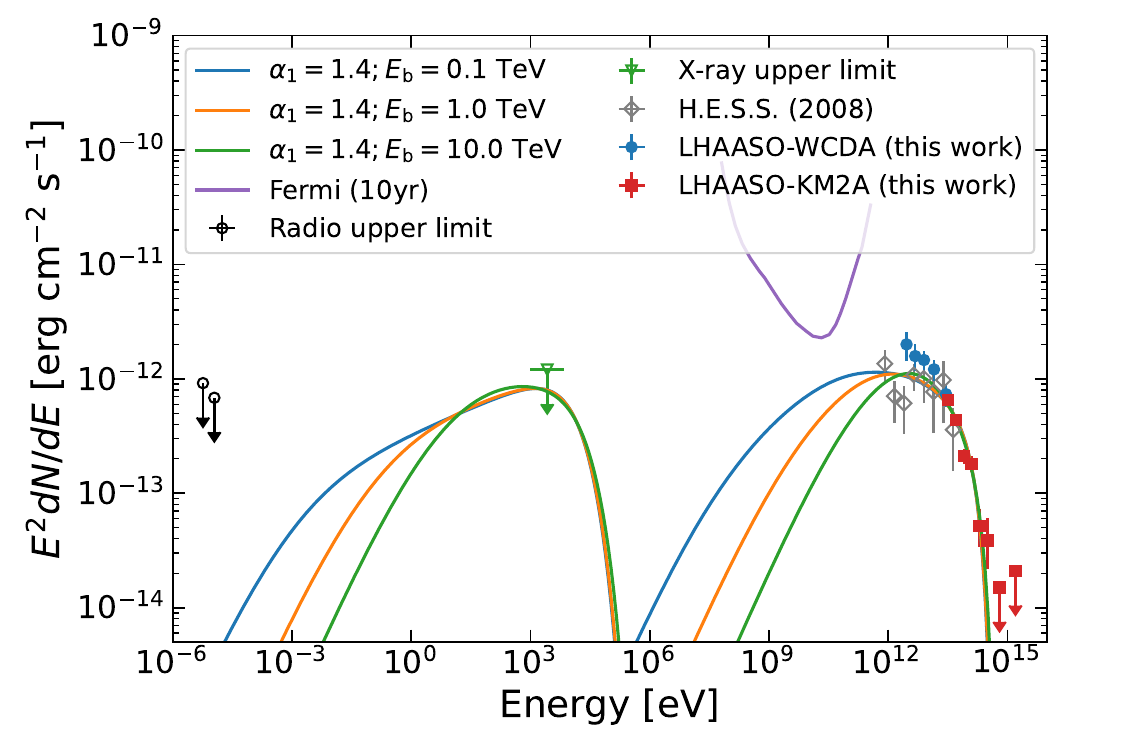}
\caption{SED for the PWN scenario.
The purple solid line shows the sensitivity of \textit{Fermi}-LAT. 
The black and green data points are the radio and X-ray upper limits obtained in Section \ref{subsec:radx}.
}
\label{fig:sed_pwn}
\end{figure}

\begin{table}
\setlength{\tabcolsep}{8pt}
\center
\caption{Parameters used in the PWN scenario.}
  \begin{tabular}{cccccc}
  \hline\hline
  $\alpha_1\,^a$ & $E_{\rm b}\,^a$ & $B\,^a$ & $\alpha_2$ & $E_{\rm e,cut}$ & $W_{\rm e}$ \\
   & (TeV) & ($\mu$G) &    & (TeV) & ($10^{47}$ erg) \\
    \hline
  1.4 & 0.1 & 3 & 2.71 & 276 & 3.2 \\
  1.4 & 1.0 & 3 & 2.77 & 284 & 0.8 \\
  1.4 & 10.0& 3 & 3.19 & 345 & 0.3 \\
     \hline
  \end{tabular}%
  \begin{tablenotes}
  \footnotesize
    \item \textbf{Notes.}
    \item $^a$ Fixed in the fitting process.
  \end{tablenotes}
\label{tab:par_pwn}%
\end{table}%

Apart from the two hadronic cases discussed above, there is another possible case, in which the TeV--PeV \grays\ arise from leptonic emission from PWNe.
PWNe are efficient accelerators for the extremely relativistic leptons (electrons and positrons), and can generate the broadband electromagnetic wave from radio to UHE {\gray}s via synchrotron and inverse Compton (IC) scattering.
As a PWN evolves, the synchrotron flux (from radio to X-rays) decreases, whereas the IC flux (GeV-TeV {\gray}s) increases until reaching a steady state value \citep{deJager2009}.
At a certain stage, PWNe can be only bright in \gray\ band and may be responsible for some unidentified TeV sources \citep{deJager2009}.
According to the Australia Telescope National Facility (ATNF; \citealt{ATNF})\footnote{\url{https://www.atnf.csiro.au/research/pulsar/psrcat/}} Pulsar Catalogue, seven known PSRs are located within 0.5$\degree$ from the LHAASO source emission centre.
As shown in Figure \ref{fig:tsmap}, all the PSRs lie outside the WCDA $r_{\rm 39}$ extension circle.
It is possible that 1LHAASO J1857+0203u is powered by an undiscovered or putative PSR.

We assume that the distribution of the evolved leptons in PWN approximately have a smooth broken PL form with a high-energy cutoff \citep{zhang20,liuPWN24}
\begin{equation}
    \frac{dN}{d\Ee} = A_{\rm e}\Ee^{-\alpha_1
    } 
    \left[ 1+\left( \frac{\Ee}{\Eb}\right)^s\right]^{\frac{\alpha_1-\alpha_2}{s}}
    {\rm exp}\left[-\left(\frac{\Ee}{E_{\rm e,cut}}\right)^{\beta}\right],
\end{equation}
where $\Ee$ is the lepton energy, $\Eb$ the break energy, $E_{\rm e,cut}$ the cutoff energy, $s$ the smooth parameter, and $\beta$ the cut-off shape parameter. Based on the results in \citet{zhang20}, $s=1$ and $\beta=3$ are adopted. $\alpha_1$ and $\alpha_2$ are the PL index below and above the break energy, respectively. $A_{\rm e}$ is the normalization parameter and is derived from the total energy in leptons with energy above 1\,GeV, $W_{\rm e}$.

Besides the cosmic microwave background, the infra-red (IR) seed photons are also included for the IC process, which depend on the position in the Galaxy.
According to the approximate calculation method presented in \cite{shibata11}, the energy density of the IR component varies from 0.4 to the maximum value $\sim$0.8 ${\rm eV\ cm^{-3}}$ when the distance increases from 1\,kpc to $\sim$7\,kpc toward the direction of the LHAASO source.
For the energy density of the IR photon field, the median value of 0.6 ${\rm eV\ cm^{-3}}$ with a temperature of 35\,K, corresponding to a distance of 3.4\,kpc, is adopted.
Because there are only H.E.S.S. and LHAASO data, the model parameters are strongly degenerated and cannot be well constrained, especially for $\alpha_1$ and $E_{\rm b}$.
For these two parameters, we take some typical values 
$\alpha_1=1.4$ and $E_{\rm b}=[0.1$, 1.0, 10]\,TeV.
Combining the radio and X-ray flux upper limits (see Section \ref{subsec:radx}),
the fitted SEDs are displayed in Figure~\ref{fig:sed_pwn}, and the corresponding parameters are listed in Table~\ref{tab:par_pwn}.

To avoid exceeding the X-ray upper limits, the average magnetic field should be less than 3\,$\mu$G.
At the same time, $\alpha_2$ and $E_{\rm e,cut}$ are acceptable according to the known \gray-bright mature PWNe ($t_{\rm psr}>$~$\sim$10\,kyr, where $t_{\rm psr}$ is the PSR age, \citealt{Zhu2018.PWN}).
The order of magnitude of the total energy in the evolved electrons at the current stage is $10^{47}$\,erg.
Considering the radiative loss, the escape process, etc, the spin-down luminosity of the putative PSR should satisfy $L_{\rm sd}>3\times10^{35}(t_{\rm psr}/10\,{\rm kyr})^{-1}(d/3.4\,\kpc)^{-2}\,{\rm erg\,s^{-1}}$.
With this condition, there are about 100 PSRs in the ATNF catalogue if the characteristic age is limited in the range from $10^4$ to $10^5$\,yr.
So it is very possible that PSRs with the above properties are not detected but power detectable PWNe in the \gray\ band.
Based on the current data, therefore, we cannot rule out the scenario in which an evolved PWN powered by a putative PSR produces the measured \grays.

\section{Conclusion} \label{sec: conclusions}
We study the TeV \gray\ emission around SNR-HIIR complex \snr\ using 796 days of LHAASO-WCDA data and 1216 days of LHAASO-KM2A data.
After subtracting the possible contamination from other sources, 1LHAASO J1857$+$0203u is detected and fitted by an extended 2D-Gaussian source template with an extension of $r_{\rm 39}\sim0.18\degree$ in the energy range of 1--25\,TeV (WCDA component), and a point-like source template at energies above 25\,TeV (KM2A component), respectively.
The PL spectral indexes of the WCDA and KM2A components are $\sim$2.46 and $\sim$3.22, respectively.
The statistical significance of the \gray\ signal reaches 11.6$\sigma$ above 100\,TeV.

We investigated three possibilities to explain the detected \gray\ emission.
In the scenario that HIIR \seast\ is a likely source of the high-energy photons, the \gray\ emission can be fitted by a PL spectrum with an exponential cutoff with a proton index of $\sim$2.0 and a cutoff energy of $\sim$450\,TeV.
Despite the absence of massive stars (OB stars), this scenario cannot be completely ruled out.
Future multi-wavelength observations are expected to discover massive stars in the HIIR.
If this is the case, \seast\ is possibly the first HII region with detected \gray\ emission up to 100\,TeV energies.
In the SNR scenario, due to the low masses of the molecular clouds, it seems unlikely that the TeV \grays\ arise from the clouds illuminated by the protons that escaped from SNR \snr.
In the scenario that an evolved PWN is powered by a putative PSR,
the spin-down luminosity of the PSR should satisfy $L_{\rm sd}>3\times10^{35}(t_{\rm psr}/10\,{\rm kyr})^{-1}(d/3.4\,\kpc)^{-2}\,{\rm erg\,s^{-1}}$.
The search for PSRs near the centre of the TeV \gray\ emissions would be helpful to explore this possibility.
Additionally, with better angular resolution,
LACT \citep{lact23}, ASTRI \citep{astri22}, and CTA \citep{cta19}
may provide clues as to whether the GeV and TeV \gray\ sources are separate sources of different origins.

\section*{Acknowledgements}
This publication makes use of data from FUGIN, FOREST Unbiased Galactic plane Imaging survey with the Nobeyama 45-m telescope, a legacy project in the Nobeyama 45-m radio telescope.
This work makes use of Very Large Array Galactic Plane Survey (VGPS) data.
The VGPS is supported by a grant from the Natural Sciences and Engineering Research Council of Canada and from the U.S. National Science Foundation.
The National Radio Astronomy Observatory is a facility of the National Science Foundation operated under cooperative agreement by Associated Universities, Inc.
This work has made use of data from the European Space Agency (ESA) mission
{\it Gaia} (\url{https://www.cosmos.esa.int/gaia}), processed by the {\it Gaia}
Data Processing and Analysis Consortium (DPAC,
\url{https://www.cosmos.esa.int/web/gaia/dpac/consortium}). 
Funding for the DPAC has been provided by national institutions, in particular the institutions participating in the {\it Gaia} Multilateral Agreement.
This work is supported in China by NSFC grants under nos.\ 12393852, 12173018, 12121003, and U1931204 and
the National Key R\&D program of China under grants  2018YFA0404201, 2018YFA0404202, 2018YFA0404203, and 2018YFA0404204,
and in Thailand by the National Science and Technology Development Agency (NSTDA) and the National Research Council of Thailand (NRCT) under the High-Potential Research Team Grant Program (N42A650868).
W.J.Z. and X.Z. thank Silvia Celli for helpful discussions.
The authors would like to thank all staff members who work at the LHAASO site 4410 m above sea level year-round to maintain the detector and keep the electrical power supply and other components of the experiment operating smoothly.
We are grateful to the Chengdu Management Committee of Tianfu New Area for their constant financial support of research with LHAASO data.

\section*{Author Contributions}
Y.C.\ initiated and organised this study; 
W.J.Z.\ performed the data analyses of the LHAASO (WCDA+KM2A), CO-line, HI, and radio observations, and analysed the optical survey data under the guidance of B.Q.C.;
X.Z.\ and Y.C.\ led the theoretical interpretation;  
W.J.Z., X.Z., and Y.C.\ prepared the manuscript.
S.Q.X.\ and S.Z.C.\ provided the pipeline for the KM2A data analysis.
S.C.H.\ and M.Z.\ provided the pipeline for the WCDA data analysis.
H.R.W. provided the cross-check.
All other authors participated in data analysis, including detector calibration, data processing, event reconstruction, data quality check, and various simulations, and provided comments on the manuscript.


\vspace{5mm}
%\facilities{HST(STIS), LHAASO (WCDA and KM2A), AAVSO, CTIO:1.3m, CTIO:1.5m,CXO}


\software{APLpy\citep{aplpy12,aplpy19},
          Astropy \citep{astropy13,astropy18},
          Naima \citep{zab15}
          }\\


\bibliography{g35}{}
\bibliographystyle{aasjournal}



\appendix

\section{Gamma-ray absorption} \label{sec:absorption}
\begin{figure*}[ht]
    \centering
    \includegraphics[width=0.5\columnwidth]{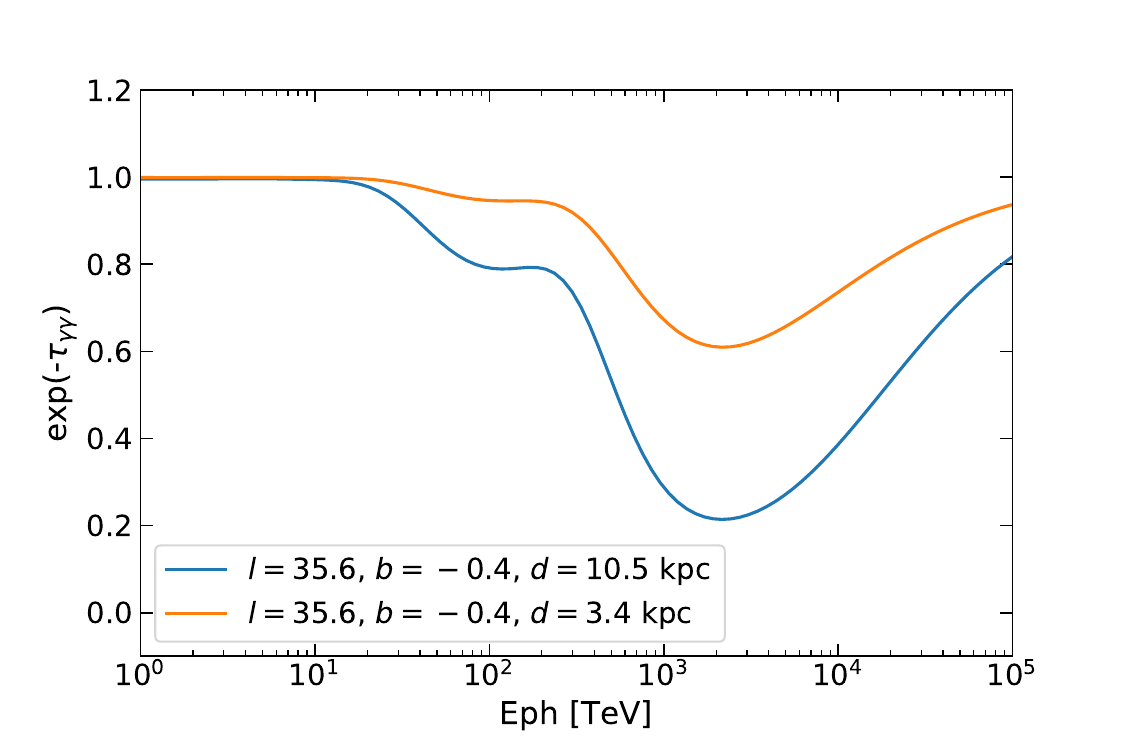}
    \caption{
    The survive probability due to the $\gamma-\gamma$ absorption.
    }
    \label{fig:gamma_absorption}
\end{figure*}

To calculate the $\gamma-\gamma$ absorption, we use the pair production cross section in \citet{gould67} and the interstellar radiation field in \citet{shibata11}. The results are displayed in Figure~\ref{fig:gamma_absorption}, showing the limited impact on the flux for $d=3.4$ kpc.
Thus we only consider the $\gamma-\gamma$ absorption in the SNR scenario in which the distance $d=10.5$~kpc is adopted.

\section{SNR scenario} \label{sec:snr_ap}
\begin{figure}[ht]
    \centering
    \includegraphics[width=0.49\columnwidth]{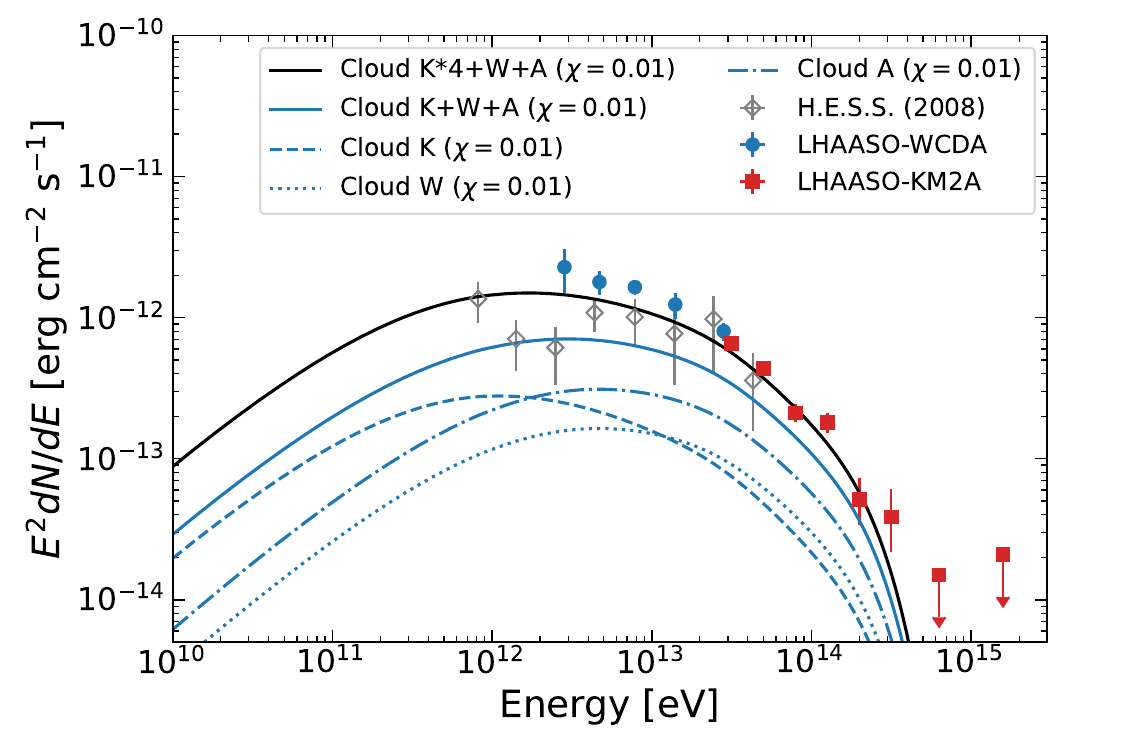}
    \includegraphics[width=0.49\columnwidth]{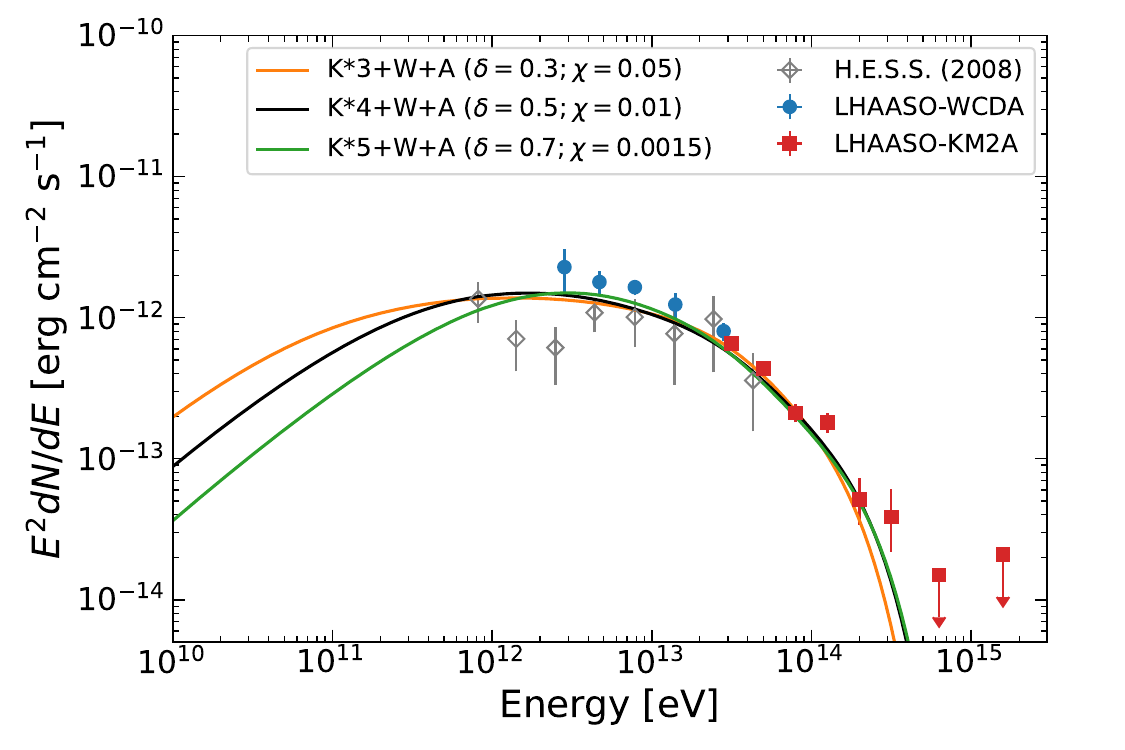}
    \caption{
    Fitted SED of the SNR scenario including the $\gamma-\gamma$ absorption.
    \textit{Left:} The blue solid line represents the total $\gamma$-ray emission from Cloud K, W, and A with $\delta=0.5$ and $\chi=0.01$. For the black solid line, the component of Cloud K is amplified by a factor of 4. 
    \textit{Right:} The $\gamma$-ray emission produced by the escaped protons is displayed in red, black, and green solid lines for $\delta$ = 0.3, 0.5, and 0.7, respectively. The detailed parameters can be found in Table~\ref{tab:sed_para_snr}.
    }
    \label{fig:sed_snr_fit}
\end{figure}

To explain the LHAASO data, the mass of clouds should be larger than the measured values. In Figure~\ref{fig:sed_snr_fit}, as an example, the mass of Cloud K is amplified by a factor of $f_{\rm K}$. The corresponding fitted parameters can be found in Table~\ref{tab:sed_para_snr}. 
\begin{table*}[ht]
\setlength{\tabcolsep}{20pt}
  \centering
  \caption{Parameters for the SNR scenario.}
    \begin{tabular}{ccccccc}
    \hline\hline
    $\eta$ & $\alpha$ & $\delta$ & $\chi$ & $s$ & $E_{\rm p,max}$ (PeV) & $f_{\rm K}^a$ \\
    \hline
    0.1 & 2.0 & 0.3 & 0.05   & 2.553 & 0.6 & 3.0 \\
    0.1 & 2.0 & 0.5 & 0.01   & 2.685 & 1.0 & 4.0 \\
    0.1 & 2.0 & 0.7 & 0.0015 & 2.968 & 3.0 & 5.0 \\
    \hline
    \end{tabular}%
    \begin{tablenotes}
        \footnotesize
        \item \textbf{Notes.}
         \item $^a$ The mass of Cloud K is amplified by a factor of $f_{\rm K}$.
    \end{tablenotes} 
  \label{tab:sed_para_snr}%
\end{table*}%

\end{document}